\documentclass[a4paper,11pt]{article}
\pdfoutput=1 
\usepackage{jheppub} % for details on the use of the package, please
\usepackage{gensymb}
\usepackage{graphicx,dcolumn,bm,epsfig,cancel,hyperref}
\usepackage{multirow}
\usepackage{amsmath}
\usepackage{amssymb, times}
\usepackage[utf8]{inputenc}
\newcommand{\comment}[1]{{}}
\numberwithin{equation}{section}
\def\beq{\begin{align}}
\def\eeq{\end{align}}
\newcommand{\bi}{\begin{itemize}}
\newcommand{\ei}{\end{itemize}}
\newcommand{\ben}{\begin{enumerate}}
\newcommand{\een}{\end{enumerate}}
\newcommand{\be}{\begin{equation}}
\newcommand{\ee}{\end{equation}}
\newcommand{\bea}{\begin{eqnarray}}
\newcommand{\eea}{\end{eqnarray}}

\usepackage{tikz}
\usepackage{subcaption}

\newcommand{\tr}{\theta_r}

\newcommand{\gev}{~\mbox{GeV}}

\newcommand{\rhos}{\rho_\phi}
\newcommand{\rhor}{\rho_r}

\newcommand{\rhom}{\rho_\chi}

\newcommand{\gam}{\Gamma_\phi}
\newcommand{\sgm}{\sigma}
\newcommand{\ds}{\delta_\phi}
\newcommand{\dr}{\delta_r}
\newcommand{\dx}{\delta_\chi}
\newcommand{\ts}{\theta_\phi}
\newcommand{\tx}{\theta_\chi}

\def\bec{\begin{center}}
\def\eec{\end{center}}
\def\beq{\begin{eqnarray}}
\def\eeq{\end{eqnarray}}
\def\fr{\frac}

\author[b,c]{Robert Wiley Deal,}
\author[a]{Kishan Sankharva,}
\author[b]{Kuver Sinha,}
\author[a]{Scott Watson}

\affiliation[a]{\footnotesize Department of Physics, Syracuse
  University, Syracuse, NY 13244, USA}
\affiliation[b]{\footnotesize Department of Physics and Astronomy, University of Oklahoma, Norman, OK, 73019, USA}
\affiliation[c]{\footnotesize Department of Physics, University of Wisconsin, Madison, WI, 53706, USA}

\emailAdd{wileydeal@wisc.edu}
\emailAdd{khsankha@syr.edu}
\emailAdd{kuversinha@ou.edu}
\emailAdd{gswatson@syr.edu}

\title{Multi-component Dark Matter and Small Scale Structure Formation}

\abstract{We consider the  evolution of non-thermal dark matter perturbations in models which contain both Weakly Interacting Massive Particles (WIMPs) and axions. Using constraints from existing observations we examine the percentage of WIMPs and axions that may comprise the cosmological dark matter budget in models with an Early Matter Dominated Epoch (EMDE) -- where entropy production is important. After carefully tracking the thermal evolution of the various species by solving the Boltzmann equations, we consider the enhancement of perturbations that may have led to early structure formation for axions and WIMPs. We   investigate the impact of enhanced perturbations on the parameter space of both species, after imposing existing constraints from indirect detection experiments.  Given these constraints we establish the feasibility of axions to form miniclusters in the early universe in EMDEs for a given percentage of allowed WIMPs. We find that EMDEs with low reheat temperatures near the BBN limit are preferred for axion minicluster formation. When the EMDE is caused by string moduli, the WIMP contribution to the relic density is set by the moduli branching to dark matter at the level of $\lesssim \mathcal{O}(10^{-4})$.}

\begin{document}
\maketitle
\flushbottom

\section{Introduction}

The Electroweak hierarchy and Strong CP problems both motivate the presence of physics Beyond the Standard Model (BSM). In the former case, constructing particle-based models that address the stabilization of the hierarchy naturally leads to the prediction of Weakly-Interacting Massive Particles (WIMPs) giving a candidate for all or part of cosmological Dark Matter (DM). Whereas, the latter problem also leads to a possible DM candidate -- the QCD axion \cite{Weinberg:1977ma,Peccei:1977hh,Wilczek:1977pj,Kim:1979if,Shifman:1979if,Zhitnitsky:1980tq,Dine:1982ah}. 

The even more difficult challenge of addressing the {\it origin} of the Electroweak hierarchy requires accounting for the presence of gravity 
in the theory. This, along with accounting for adequate cosmic inflation in the early universe typically implies additional scalar degrees of freedom and additional axion-like particles (ALPs). Some of these fields can become populated following inflation and dominate the energy density resulting in a departure from a strictly thermal history prior to Big Bang Nucleosynthesis (BBN). 

Originally posed as the cosmological moduli problem \cite{Hall:1983iz, deCarlos:1993wie, Coughlan:1983ci, Banks:1993en}, in recent years these Early Matter Dominated Epochs (EMDEs) have been viewed more as an opportunity to probe the early history of the Universe and make connections to string phenomenology. The primordial origin, abundance, and clustering of WIMPs and axions can be altered by the pre-BBN cosmic evolution. Moreover, when these particles decay they can lead to significant entropy production altering the primordial abundance of WIMPs and axions. EMDEs have been extensively studied in recent years in the context of their effect on DM searches (for recent reviews, we refer to \cite{Allahverdi:2020bys,Kane:2015jia}; for earlier studies, we refer to \cite{Giudice:2000ex, Gelmini:2006pw}) and their connections to the physics of moduli stabilization \cite{Allahverdi:2013noa,Allahverdi:2014ppa,Cicoli:2022uqa,Dutta:2009uf,Acharya:2008bk,Acharya:2009zt}.

Much of the previous work on EMDEs has focused on the effects of such eras on dark sectors consisting of a single component, often a standard WIMP. For example, the effects on DM indirect and direct detection \cite{Blanco:2019eij, Easther:2013nga}, connections with baryogenesis \cite{Kane:2011ih, Allahverdi:2010rh, Allahverdi:2010im, Allahverdi:2012gk} and structure formation \cite{Miller:2019pss,Delos:2018ueo,Delos:2017thv,Fan:2014zua} have been probed in detail, yielding a rich phenomenological probe of the early Universe. 

The purpose of this paper is to initiate a study of EMDEs for \textit{multicomponent} DM, with an emphasis on the possibility of DM substructure formation. The focus on enhanced DM sub-structure formation comes from the fact that this is one of the most important and universal effects of EMDEs: subhorizon DM perturbations grow linearly during EMDEs and can lead to an enhancement in the growth of sub-structure on small scales. For a dark sector model with multiple interacting species or a single self-interacting one, a study combining alternative cosmological histories with structure formation can be expected to be quite challenging. As a first step, it is therefore reasonable to restrict ourselves to models where the different components contributing to the relic density do not interact with each other. Given the fact that EMDEs typically arise in string compactifications with both scalar moduli as well as their pseudoscalar axion-like partners,  it is natural to consider the multicomponent sector to be comprised of WIMPs and axions, specifically the QCD axion. The phenomenology of such scenaros has been studied in detail, especially in a supersymmetric setting \cite{Baer:2020kwz, Baer:2019xww, Bae:2013hma, Baer:2011uz, Bae:2014rfa}.

Given a multi-component DM sector consisting of WIMPs and axions, one can focus on the question of sub-structure formation in either sector, and ask whether the parameter space for such sub-structure formation is consistent with the observed relic density and various observational bounds. For concreteness, we will focus on the question of sub-structure formation in the axion sector (axion miniclusters)\footnote{Sub-structure formation in the WIMP sector has been studied in detail by a subset of the current authors \cite{Fan:2014zua}.} and study how the existence of such miniclusters  affects the properties and thermal  history of the WIMP sector. EMDEs  are expected to play a critical role in the formation of axion miniclusters in the case where Peccei-Quinn (PQ) symmetry breaking occurs before inflation, which is the case we study. We first carefully track the thermal history of the modulus, axions, WIMPs and radiation by solving the coupled Boltzmann equations, and then study the evolution of the perturbation of modes that enter the Hubble radius during the EMDE and subsequently through reheating. Imposing the condition for the formation of miniclusters leads to constraints on the decay temperature of the modulus, which in turn impact the parameter space available for the WIMP sector. In particular, we study the resulting constraints on WIMPs coming from indirect detection data.

The question of axion minicluster formation with EMDEs has garnered some attention in recent years \cite{Nelson:2018via, Blinov:2019jqc, Visinelli:2018wza}. Our work differs from these studies mainly by the introduction of the second component of DM (WIMPs) and by a careful treatment of the interplay between the components and their thermal histories. As we have argued, string theory naturally comes with situations where moduli and axions co-exist, and WIMPs are  motivated by a solution to the hierarchy problem. But one can also ask: if one is interested solely in axion sub-structure with EMDEs, is one \textit{forced} to introduce a second DM component? The answer we give is to the affirmative, at least for a large range of values of the axion decay constant.  A careful computation of the interplay between the axion contribution to the relic density  and the possibility of minicluster formation  reveals that sub-structure formation prefers low modulus reheat temperatures $T^\phi_D \lesssim \mathcal{O}(50)$ MeV, which can only accommodate pure axion DM for decay constants $f_a \gtrsim 10^{15}$ GeV.  Smaller values of the decay constant will require a second DM component. These results are depicted in Fig.~\ref{fig:axionDM_untuned}, and we find that while tuning the initial misalignment can relax these requirements somewhat, the qualitative trends remain unaltered (a mildly tuned case is depicted in Fig.~\ref{fig:axionDM_tuned}). The preference of low modulus decay temperatures for minicluster formation is due to the fact that a prolonged EMDE is required for perturbations to grow to the requisite level. And it is precisely in the low reheat regime that axion decay constants below the GUT scale of $\sim 10^{15}$ GeV will necessitate a second DM component.

In the low reheat regime to which we are forced, the WIMPs have to be produced from the modulus with small branching ratios $\lesssim \mathcal{O}(10^{-4})$, corresponding to the ``branching scenario" \cite{Allahverdi:2020bys,Kane:2015jia}. For larger branching ratios, WIMP re-annihilation after production from modulus decay becomes the predominant mechanism for setting the relic density (the ``annihilation scenario") and such cases are highly constrained by indirect detection data. Whether the small branching ratios of moduli into WIMPs required for the branching scenario to be successful can be obtained from actual string compactificatons is a question we leave for the future.

%In this paper, we consider how an Early Matter Dominated Epoch (EMDE) prior to BBN affects the growth of DM sub-structure and cosmological observations.  and early Universe cosmology

%In EMDE it complicates the way to interpret cosmological predictions prior to BBN. Other questions arise -- what is the true nature of dark matter? How did the universe evolve prior to BBN? How did Peccei - Quinn (PQ) breaking occur during or after inflation? We attempt to address these questions aiming to make them more precise with the most recent data. 

%Former EMDE studies have taught us that the way structure growth led to inhomogeneity in the (smoothed by inflation) axion field can alter predictions for the critical temperatures of importance and their eventual composition for structure. In this paper, we are investigating two simultaneous DM candidates with the latest observations. Can ALPs form mini-clusters given the presence of WIMPS? We try to address this in what follows. 

The paper is structured as follows. In Section \ref{nonthbg} we introduce EMDEs and axions, set our notation, and solve the Boltzmann equations for the various components. Section \ref{perturb} addresses substructure formation in the presence of both axions and WIMPs. We then address the formation of axions in the formation of miniclusters and then give our results for WIMP constraints in Secction \ref{sec:minicluster}. We then conclude.

\section{Non-thermal dark matter production} \label{nonthbg}
In this section, we discuss the production of both axion and WIMP dark matter in the context of EMDEs. We begin with a brief review of the QCD axion and its background evolution, as well as its connections to inflation.
We then present the zeroth-order Boltzmann equations which govern the evolution of the densities of WIMPs and axions, in addition to radiation and the modulus.
This allows us to analyze the qualitative behavior of the thermal evolution of the different species and estimate the relic densities in limiting cases.
Finally, we detail our procedure for obtaining numerical solutions of the Boltzmann equations which we utilize in Sec.~\ref{sec:minicluster} and Sec.~\ref{sec:WIMPs}.

\subsection{The QCD Axion}

The QCD axion is a pseudo Nambu-Goldstone boson generated by spontaneous breaking of the Peccei-Quinn (PQ) symmetry which involves a complex scalar field called the PQ field. The Lagrangian for the PQ field $\xi$ is given by
\be\label{eq:PQ_Lagrangian}
\mathcal{L}_{\xi} = \frac{1}{2}\partial_{\mu}\xi^{\dagger} \, \partial^{\mu}\xi - \frac{\lambda}{4}(\xi^{\dagger}\xi - f_a^2)^2,
\ee
where $f_a$ is the vacuum expectation value of the PQ field. The expectation value of the $\xi$ can be written as $\langle\xi(x)\rangle = f_ae^{i\varphi(x)/f_a}$ where the phase field $\varphi(x)$ is identified as the axion field. Here, the axion field $\varphi(x)$ does not have any potential due to PQ symmetry and therefore, it can take any value. However, after PQ symmetry breaking, the axion receives a potential via its coupling to gluons. This potential is minimized at a strong CP conserving value. Since before PQ transition the axion field takes on any random value, after the transition its initial value may be `misaligned' from the minima of its potential. The energy stored in this misaligned axion later takes the form of dark matter.

The axion mass is expected to be temperature-dependent due to finite-temperature contributions from instantons, and takes the schematic form\footnote{
Here, we have set the domain wall number $N_{\text{DW}}=1$ for simplicity.  This will not have significant impact on our results since most axion models have $N_{\text{DW}} \sim \mathcal{O}(1-10)$.
} \cite{Hertzberg:2008wr,Gross:1980br}
\begin{equation}
    \label{eq:axionMass}
    m_a(T)
    =
    \left(
        6.2
        \times 
        10^{-3}
        \text{ GeV}
    \right)
    \left(
        \frac{
            1 \text{ GeV}
        }{
            f_a
        }
    \right)
    \times
    \begin{cases}
        1
        &
        (
            T
            \lesssim 
            \Lambda_{\text{QCD}}
        )
        \\
        b
        \left(
            \frac{
                \Lambda_{\text{QCD}}
            }{
                T
            }
        \right)^4
        &
        (
            T
            \gtrsim 
            \Lambda_{\text{QCD}}
        )
    \end{cases}
\end{equation}
where $b=0.018$ and $\Lambda_{\text{QCD}} \sim 200$ MeV.
The axion potential after PQ symmetry breaking can then be written approximately as
\be\label{eq:axion_potential}
V_a(x, T) = f_a^2m_a^2(T)\bigg[1 - \cos\bigg(\frac{\varphi(x)}{f_a}\bigg)\bigg] \simeq \frac{1}{2}m_a^2(T)\varphi^2(x)
\ee
where $m_a(T)$ is given by Eq.~\eqref{eq:axionMass}.

The background equation of motion for the axion field in an expanding universe is thus given by
\be\label{eq:axion_eom}
\Ddot{\varphi} + 3 H(t) \, \Dot{\varphi} + m_a^2(T)\varphi = 0
\ee
where $H(t)$ is the Hubble scale, and we use overdots to represent derivatives with respect to comoving time $t$. 
Since the axion is minimally coupled, the above equation can be used to obtain the evolution of the axion energy density, irrespective of other components in the universe. 
However, solving this equation is non-trivial for several reasons. First, note that Eq.~\eqref{eq:axion_eom} is an equation of motion for a damped harmonic oscillator. However, both its damping coefficient and its mass are effectively time-dependent. 
This makes a general analytical solution implausible. 
Second, the axion oscillates very rapidly once the damping coefficient ($\sim H$) is smaller than the axion mass ($m_a$). 
This makes Eq.~\eqref{eq:axion_eom} a stiff differential equation which is difficult to solve numerically. Fortunately, we can solve this equation approximately and deduce the general trend of the evolution of the axion energy density.

Note that the second term, $3H\Dot{\varphi}$, behaves like a time-dependent friction term which does not allow the axion field to oscillate until $H \simeq m_a$. 
This gives the condition for critical time $t_{\rm osc}$ at which the axion oscillations begin:
\be\label{eq:axion_critical_time}
H(t_{\rm osc}) \simeq m_a(T(t_{\rm osc})).
\ee
We can thus take the axion field to be frozen for $t < t_{\rm osc}$, which produces an equation of state $w_a = -1$.
The axion field then oscillates very rapidly for $t > t_{\rm osc}$ such that its average equation of state in this regime is $\langle w_a \rangle = 0$. 
In other words, the axion field behaves similarly to a cosmological constant before $t_{\rm osc}$ with its number density $n_a =$ constant\footnote{
    Due to the temperature dependence of the axion mass, the axion energy density while it is frozen should actually \textit{increase} as the universe expands due to its increasing mass.
}, and it behaves like dark matter after $t_{\rm osc}$ where its number density dilutes as $n_a \sim a^{-3}$ where we use $a$ to denote the scale factor. Thus, we can estimate the number density of axions as
\be\label{eq:axion_number_density}
n_a(t) \simeq \frac{1}{2}m_a(t_{\rm osc})
\, f_a^2\theta_i^2 \left(
\frac{a(t_{\rm osc})}{a(t)}
\right)^3,
\ee
where $\theta_i = \varphi_0/f_a \in [0, \pi)$ is the initial misalignment angle, which is random and is typically assumed to be $\mathcal{O}(1)$. Now, the axion energy density at time $t$ is simply given by $\rho_a(t) = m_a(t)n_a(t)$.

It is worth mentioning that much of this discussion can also apply to the modulus as well.
During the inflationary epoch, moduli are expected to be displaced from their minima \cite{Dine:1995kz,Dine:1995uk} and frozen in place by Hubble friction.
Once the universe has expanded sufficiently (i.e. when $H \sim m_\phi$), they begin to oscillate and behave as cold matter.
However, moduli typically are not protected by such symmetries and thus are expected to have a Planck-scale initial amplitude \cite{Dine:1995kz,Dine:1995uk,Cicoli:2016olq}.
Additionally, moduli are typically assumed to be very massive with $m_\phi \gtrsim \mathcal{O}(10-50 \text{ TeV})$ to avoid conflicts with BBN.  

\subsection{PQ Breaking Scale and Inflation}

The relationship between the PQ breaking scale and the scale of inflation has been  studied both in the context of standard as well as modified cosmologies \cite{Hertzberg:2008wr, Poulin:2018dzj,Visinelli:2009kt, Baer:2011uz,  Visinelli:2018wza, Hlozek:2017zzf}.  Denoting the Hubble scale during inflation by $H_I$, the two scenarios are  $f_a < H_I/2\pi$ (axion absent during inflation) and $f_a > H_I/2\pi$ (axion present during inflation). For the former case, PQ symmetry breaking occurs after inflation and the initial misalignment angle $\theta_i$ acquires different values within the same Hubble patch. The formation of axion miniclusters in this case has been studied extensively \cite{Kolb:1993zz,Hogan:1988mp,Turner:1983sj} and will not be probed further in the current work.

For the second case, PQ symmetry breaking occurs before inflation, and the initial misalignment angle takes only a single value in the Hubble horizon.  In this case, axion isocurvature fluctuations are constrained by the cosmic microwave background and BBN. For standard cosmological histories, the constraint can be expressed as a relation between $H_I, \theta_i,$ and $f_a$ as follows \cite{Visinelli:2009zm,Planck:2018jri}:
\be
H_I \, \lesssim \, 10^{-5} \, \left(\frac{\Omega_{\rm CDM}}{\Omega_a}\right) \theta_i f_a \,\,.
\ee
Constraints coming from isocurvature fluctuations in the case of scenarios with decaying scalar fields were worked out in \cite{Visinelli:2009kt}. In this case, one has
\be
H_I \, \lesssim \, 1.0 \times 10^{8} \left(\frac{\Omega_{\rm CDM}}{\Omega_a}\right) \left(\frac{f_a}{10^{12}\, {\rm GeV}}\right) \,\,{\rm GeV} \,\,,
\ee
which holds for $f_a < 10^{15} \left(\frac{T^{\phi}_D}{1 {\rm MeV}}\right)^{-4/3}$  GeV. 
For higher values of $f_a \gtrsim  10^{15} \left(\frac{T^{\phi}_D}{1 {\rm MeV}}\right)^{2}$ GeV, one has 
\be
H_I \, \lesssim \, 5.0 \times 10^{10} \left(\frac{\Omega_{\rm CDM}}{\Omega_a}\right) \left(\frac{T^{\phi}_D}{1 {\rm MeV}}\right)^{-1/2} \,\,{\rm GeV} \,\,.
\ee
For details of other regimes of $f_a$, we refer to \cite{Visinelli:2009kt}. Independent of axion physics, the scale of inflation is constrained by Planck \cite{Planck:2018jri} to be 
\be
H_I \, \lesssim \, 2.5 \times 10^{-5} \, m_P \,\,.
\ee

For the remainder of this work, we will be agnostic about the details of the inflationary sector, and assume a value of $H_I$ such that all constraints are satisfied.

\subsection{The Boltzmann equations}

We now shift to the discussion of the zeroth-order Boltzmann equations to study the production of WIMP and axion dark matter in modulus-dominated non-thermal cosmologies.
We begin with the Friedman, Lemaitre, Robertson, Walker (FLRW) metric
\be
ds^2= -dt^2 +a^2(t)d\vec{x}^2,
\ee
with the resulting Hubble equation
\bea
3 H^2 m_p^2 &=& \sum_\alpha \rho_{(\alpha)}, \label{FRW0}
\eea
where $\alpha \in \{ \phi,\chi,r,a \}$ represents moduli, WIMPs, radiation, and axions, respectively.
We work in units of the reduced Planck mass $m_p=2.44 \times 10^{18} \gev$.  

For this work, we adopt a coupled Boltzmann equation approach which includes tracking the particles with all other degrees of freedom treated as radiation.
Instead of working with comoving time, it is convenient to express the full set of equations as a function of e-foldings, $H dt= dN=d(\ln{a})$, producing the set of equations
\bea \label{back}
\label{rhos}\fr{d\rhos}{dN}  &=& -3(1+w_\phi)\rhos - \fr{\gam}{H}\rhos, \\
\label{rhm}\fr{d\rhom}{dN} &=&-3(1+w_\chi)\rhom - \frac{\langle \sgm v \rangle_\chi }{m_\chi H} \left(\rhom^2 - \overline{\rho}_\chi^2 \right) +B_{\phi \rightarrow \chi }\left( \fr{\gam}{H}\right) \rhos, \\
\label{axe} \frac{dn_a}{dN} &=& -3\left( 1+w_a \right) n_a,\\
\label{rhr}\fr{d\rhor}{dN} &=& -4\rhor  +\frac{\langle \sigma v \rangle_\chi}{m_\chi H}\left(\rhom^2 - \overline{\rho}_\chi^2 \right) +(1-B_{\phi \rightarrow \chi }) \fr{\gam}{H} \rhos,
\eea
subject to the energy constraint Eq.~\eqref{FRW0} and where we have kept the equation of state $w_\alpha$ general. 
Note that we have also expressed the equation governing the evolution of the axion in terms of its number density, $n_a$, which is simpler to work with than its energy density.\footnote{
    The Boltzmann equation governing the evolution of the axion's energy density requires the inclusion of a term of the form $d m_a / dN$ due to its aforementioned temperature-dependent mass.
}
Indeed, because the axion field is always non-relativistic we can always retrieve its energy density from knowledge of its number density and its mass.
Before we continue, it is worth discussing the limiting cases of the above Boltzmann equations which reduce to familiar expressions typically used in the literature.

\subsection{Qualitative behavior and limiting cases}

Our first task is to estimate the scale of modulus decay.
From the Boltzmann equation of the modulus, Eq.~\eqref{rhos}, we see the decay term becomes sizeable when 
\begin{equation}
    \Gamma_\phi 
    \sim 
    H
    .
\end{equation}
In the sudden decay approximation and assuming a radiation-dominated universe after the modulus decays, we can estimate the decay temperature $T_D^\phi$ as 
\begin{equation}
    T_D^\phi 
    \simeq
    \sqrt{
        m_P
        \Gamma_\phi
    }
    \left( 
        \frac{
            90
        }{
            \pi^2
            g_*(T_D^\phi)
        }
    \right)^{1/4}
    .
\end{equation}
This temperature is also frequently referred to as the ``reheating temperature'' in much of the literature.
As moduli are expected to be only gravitationally-coupled, we note that the decay temperature can be expected to be quite low.  
If $T_D^\phi \lesssim \mathcal{O}(1 \text{ MeV})$, decay occurs during or after Big Bang Nucleosynthesis (BBN) \cite{Kawasaki:1999na, Kawasaki:2000en, Hasegawa:2019jsa} and - due to the substantial entropy production - erases the successful predictions of BBN.  
This is the standard cosmological moduli problem (CMP) and sets a lower bound on the decay width $\Gamma_\phi \gtrsim 5 \times 10^{-25}$ GeV, which in turn sets a constraint on the modulus mass for a particular set of couplings.
As a precise limit on the modulus decay temperature is model-dependent \cite{Hasegawa:2019jsa}, we adopt a naive BBN cutoff of 1 MeV throughout.

It is also useful to estimate the scale at which entropy begins to be injected into the thermal bath.
To proceed, we first define the temperature at which the modulus begins to dominate the energy density of the universe as $T_e^\phi$.
From imposing $\rho_r \sim \rho_\phi$ and assuming a radiation-dominated universe before modulus domination, the temperature of modulus-radiation equality can be shown to be\footnote{
    This form assumes that modulus oscillations onset only after any inflationary reheating has finished so that the modulus is unaffected by entropy production from the decay of the inflaton.  See e.g. \cite{Bae:2022okh, WileyDeal:2023sry} for details.
} \cite{Bae:2022okh}
\begin{equation}
    T_e^\phi 
    \simeq 
    \frac{3}{2}
    \left( 
        \frac{10}{\pi^2 g_*(T_{\text{osc}}^\phi)}
    \right)^{1/4}
    \left(
        \frac{\phi_0}{m_P}
    \right)^2
    \sqrt{m_\phi m_P}
\end{equation}
where $\phi_0$ is the initial amplitude of the modulus.
As the universe is matter-dominated between $T_e^\phi$ and $T_D^\phi$, it is a simple matter to relate the Hubble scales at both eras.
Upon noting that entropy is conserved between $T_e^\phi$ and - by definition - the entropy injection temperature $T_S^\phi$, while entropy is \textit{not} conserved between $T_S^\phi$ and $T_D^\phi$, we can arrive at the following estimate for the entropy injection temperature:
\begin{equation}
    T_S^\phi
    \simeq
    \left(
        T_e^\phi
        \left(
            T_D^\phi
        \right)^4
    \right)^{1/5}
    .
\end{equation}
Although this entropy injection temperature will not be necessary for our discussion on WIMP production, it will become a key but implicit feature in determining the axion relic density within a modulus-dominated cosmology.

We now consider the limiting cases for WIMP production, beginning with the case where WIMP annihilations are dominant.
Considering the regime close to (or below) the thermal freeze-out scale and assuming the WIMPs behave as cold dark matter in this regime (i.e. $w_\chi = 0$ and $\rho_\chi \simeq m_\chi n_\chi$), the Boltzmann equation for WIMPs can be rewritten in terms of the abundance yield $Y_\chi \equiv n_\chi / s$ as
\begin{equation}
    \frac{
        d Y_\chi
    }{
        dN
    }
    \simeq
    -
    \frac{\langle \sigma v \rangle_\chi}{H}
    Y_\chi^2
    s
    .
\end{equation}
Taking the limit of constant $\langle \sigma v \rangle_\chi$, this equation can be straightforwardly integrated from the scale of modulus decay $T_D^\phi$ to the present time to give the result 
\begin{equation}
    Y_\chi 
    \simeq
    \frac{
        H(T_D^\phi)
    }{
        \langle \sigma v \rangle_\chi
        \,
        s(T_D^\phi)
    }
    .
\end{equation}
Assuming no significant source of entropy production after the modulus has decayed, the WIMP relic density can be estimated by \cite{Kolb:1990vq}
\begin{equation}
    \label{eq:relicDensityEstimate}
    \Omega_\chi h^2
    \simeq
    \frac{
        m_\chi
        Y_\chi
        h^2
    }{
        \rho_c / s_0
    }
\end{equation}
where $s_0 \sim 2970 \text{ cm}^{-3}$ and $\rho_c / h^2 \sim 1.05 \times 10^{-5} \text{ GeV}\cdot \text{cm}^{-3}$.
This can also be used to relate the produced WIMP relic density $\Omega_\chi h^2$ to the expected WIMP relic density produced in a standard thermal scenario $\Omega_\chi^{\text{th}} h^2$:
\begin{equation}\label{relicannnont}
    \Omega_\chi h^2
    =
    \Omega_\chi^{\text{th}}
    h^2
    \times
    \max
    \{ 
        T_f/T_D^\phi,
        1
    \}
    .
\end{equation}
This well-known result is typically coined the \textbf{annihilation scenario} in the literature \cite{Allahverdi:2013noa,Allahverdi:2014ppa,Cicoli:2022uqa}.

The other limiting case for the production of WIMPs is when the modulus decay term dominates. 
Making again the assumption that the WIMPs behave as cold dark matter, the Boltzmann equation for WIMPs can be written as 
\begin{equation}
    \frac{
        d Y_\chi
    }{
        dN
    }
    =
    B_{\phi \rightarrow \chi}
    \frac{
        \Gamma_\phi
    }{
        H
    }
    Y_\phi
\end{equation}
where we again rewrite the Boltzmann equation in terms of the abundance yield $Y_\chi$.
Integrating this equation from the scale of modulus decay to the present time then gives the result 
\begin{equation}\label{branchingyield}
    Y_\chi 
    \simeq
    B_{\phi \rightarrow \chi}
    Y_\phi(T_D^\phi).
\end{equation}
Once again, the WIMP relic density can be estimated with Eq.~\eqref{eq:relicDensityEstimate}.
This result is typically denoted as the \textbf{branching scenario} \cite{Allahverdi:2013noa,Allahverdi:2014ppa,Cicoli:2022uqa}.

We now turn to a discussion of the axion.
In this work, we concern ourselves only with the case where the modulus decays \textit{after} the axion field has begun to oscillate, i.e. $T_D^\phi < T_{\text{osc}}^a$, so that axion miniclusters might form if PQ symmetry breaking occurs at or above the inflationary scale \cite{Nelson:2018via}.
This has the crucial property that entropy dilution from the modulus will reduce the late-time relic density of the misalignment axions \cite{Baer:2023bbn}.
In a ``decay-dominated'' background - i.e. during the era in which the modulus injects significant entropy into the thermal bath - we can then estimate the Hubble scale at the onset of axion oscillations in terms of the modulus decay temperature as \cite{Kolb:1990vq}
\begin{equation}
    \label{eq:axionOscillationHubble}
    H(T_{\text{osc}}^a)
    \simeq
    \frac{
        g_*(T_{\text{osc}}^a)  
    }{
        g_*(T_{D}^\phi)
    }
    \frac{
        (T_{\text{osc}}^a)^4
    }{
        (T_{D}^\phi)^2
    }
    \sqrt{
        \frac{
            \pi^2
            g_*(T_D^\phi)
        }{
            90 m_P^2
        }
    }
\end{equation}
where we assume a transition to a radiation-dominated universe at $T_D^\phi$.
It is then straightforward to compute $T_{\text{osc}}^a$ from Eq.~\eqref{eq:axionMass} and Eq.~\eqref{eq:axion_critical_time}, which we find to be roughly $T_{\text{osc}}^a \sim \mathcal{O}(0.1 - 1 \text{ GeV})$ depending on the value of $T_D^\phi$.
Having found the axion oscillation temperature, we can then estimate the axion number density when oscillations begin by 
\begin{equation}
    \label{eq:axionInitialCondition}
    n_a(T_{\text{osc}}^a)
    \simeq
    \frac{1}{2}
    m_a(T_{\text{osc}}^a)
    \mathcal{A}_0^2
\end{equation}
where, accounting for anharmonic effects, the initial oscillation amplitude $\mathcal{A}_0$ is given by \cite{Visinelli:2009zm}
\begin{equation}
    \label{eq:axionAmplitude}
    \mathcal{A}_0^2
    =
    1.44
    \,
    f_a^2
    \theta_i^2
    \,
    \left[ 
        \log 
        \left( 
            \frac{
                e
            }{
                1-\theta_i^2/\pi^2
            }
        \right)
    \right]^{7/6}
\end{equation}
in terms of the Peccei-Quinn scale $f_a$ and the initial misalignment angle $\theta_i$.
Accounting for the remaining matter-dominated period between $T_{\text{osc}}^a$ and $T_D^\phi$ so that 
\begin{equation}
    n_a(T_D^\phi)
    =
    n_a(T_{\text{osc}}^a)
    \left(
        \frac{
            a(T_{\text{osc}}^a)
        }{
            a(T_D^\phi)    
        }
    \right)^3
    \simeq
    n_a(T_{\text{osc}}^a)
    \left( 
        \frac{
        g_*(T_D^\phi)(T_D^\phi)^4 
        }{
        g_*(T_{\text{osc}}^a)(T_{\text{osc}}^a)^4 
        }
    \right)^2
\end{equation}
and assuming no entropy production after $T_D^\phi$, we arrive at the relic density estimate for misalignment axions:
\begin{equation}
    \Omega_a
    h^2
    \simeq
    \frac{
        45
    }{
        4
        \pi^2
    }
    \frac{
        m_a(T=0)
        \,
        m_a(T_{\text{osc}}^a)
        \,
        \mathcal{A}_0^2
    }{
        \rho_c / (h^2 s_0)
    }
    \frac{
        g_*(T_D^\phi)
    }{
        g_*^2(T_{\text{osc}}^a)
    }    
    \frac{
        (T_D^\phi)^5    
    }{
        (T_{\text{osc}}^a)^8
    }
    .
\end{equation}

\subsection{Numerical solutions to Boltzmann equations}

Now that we have a few qualitative estimates for the scales and predicted abundances, we can return to the procedure of numerically integrating Eqs.~\eqref{rhos}-\eqref{rhr}.
A small codebase was written in Python incorporating these differential equations, which were solved using a backward differentiation formula (BDF) implementation in SciPy's ode class \cite{2020SciPy-NMeth}.
At each step, we recover the temperature from the radiation energy density.
It is then a simple matter to recover the current step's Hubble parameter and the axion mass, as well as the equilibrium energy density of the WIMP which, making the conventional definition $x \equiv m_\chi / T_\chi$, can be approximated as \cite{Kolb:1990vq,Baer:2011uz}
\begin{equation}
        \overline{\rho}_\chi 
        =
        \begin{cases}
            \frac{7}{8} 
            \frac{\pi^2}{30} 
            \,
            gT^4
            &
            (
                3 / 2 \leq x^{-1}
            )
            \\
            \frac{
                1
            }{
                    2 \pi^2
            }
            \,
            g 
            T^4
            \,
            x^2
            \left(
                x
                K_1( x )
                +
                3
                K_2( x )
            \right)
            &
            (
                1 / 10 < x^{-1} < 3 / 2
            )
            \\
            \frac{
                1
            }{
                (2 \pi )^{3/2}
            }
            \,
            g
            T^4 
            \,
            x^{5/2} 
            \exp( - x )
            &
            (
                x^{-1} \leq 1 / 10
            )
        \end{cases}
\end{equation}
where the middle case interpolates between the relativistic and non-relativistic limits, and where $g=2$ for a spin-$1/2$ WIMP.

Within the Boltzmann equation implementation, our code adopts the equation of state
\begin{equation}
    w_\alpha
    =
    \begin{cases}
        -1
        &
        (
            m_\alpha < H
        )
        \\
        0
        &
        (
            m_\alpha \geq H
        )
    \end{cases}
\end{equation}
for both the modulus and axion (i.e. $\alpha \in \{ \phi, a \}$), where the axion uses its mass for the current step.
The equation of state for WIMPs, however, is slightly more complicated as we numerically integrate from the regime where WIMPs are relativistic to the regime where they become cold.  
The WIMP equation of state can be approximated as \cite{Kolb:1990vq,Baer:2011uz}
\begin{equation}
    w_\chi 
    =
    \begin{cases}
        1/3
        &
        (
            3 / 2 \leq x^{-1}
        )
        \\  
        \left(
            x
            \frac{
                K_1(x)
            }{
                K_2(x) 
            }
            +
            3
        \right)^{-1}
        &
        (
            1 / 10 < x^{-1} < 3 / 2
        )
        \\
        0
        &
        (
            x^{-1} \leq 1 / 10
        )
    \end{cases}
\end{equation}
where we again use $x \equiv m_\chi / T_\chi$.
However, there is an additional complication - when WIMPs are produced via modulus decay, they may be highly relativistic even at temperatures below the standard thermal freeze-out temperature simply due to the comparatively large modulus mass \cite{WileyDeal:2023sry}.  
Thus, the WIMP equation of state requires a subtle distinction in its temperature.
In this work, we accomplish this by incorporating an additional Boltzmann equation governing the number density of the WIMP.
This additional equation is entirely independent of the equation of state, which allows us to both numerically fit $T_\chi$ through a comparison of $\rho_\chi$ to $n_\chi m_\chi$ as well as provide a secondary estimate for the WIMP relic density.
Interestingly, we find agreement between the number density and energy density Boltzmann equations to well below the percent-level for the calculated WIMP relic density.

All that remains is to determine the initial conditions.
As the results are expected to be independent of the inflationary reheating temperature\footnote{
This behavior was confirmed in \cite{Baer:2023bbn} for predictions of both WIMP and axion dark matter, as well as production of dark radiation, using a similar codebase.
} $T_R$, we set our initial conditions at this point.  
Assuming that the universe is radiation-dominated at $T_R$ (with a period of EMD driven by the modulus occurring shortly after), the initial conditions follow immediately.
The initial radiation energy density is set simply by 
\begin{equation}
    \rho_r^{(0)}
    =
    \frac{
        \pi^2
    }{
        30
    }
    g_*(T_R)
    T_R^4
    .
\end{equation}
The initial energy density of the modulus is given by
\begin{equation}
    \rho_\phi^{(0)}
    =
    \frac{1}{2}
    m_\phi^2
    \phi_0^2
\end{equation}
where we take $\phi_0 = m_P$, as expected on dimensional grounds \cite{Dine:1995kz,Dine:1995uk} and was found in an explicit calculation for K\"{a}hler inflation \cite{Cicoli:2016olq}.
We then take Eq.~\eqref{eq:axionInitialCondition} for the initial number density of the axion - which is updated at each step until oscillations begin to account for the varying mass.
Finally, a quick comparison of the interaction rate 
$ 
\langle \sigma v \rangle_\chi 
\overline{\rho}_\chi(T_R) 
/ 
m_\chi 
$
to the Hubble rate $H(T_R)$ for the WIMP cross sections and masses of interest suggest that the WIMPs begin in thermal equilibrium.
Thus, we take $\rho_\chi^{(0)} = \overline{\rho}_\chi(T_R)$.

\begin{figure}[htb!]
    \centering
    \includegraphics[scale=0.8]{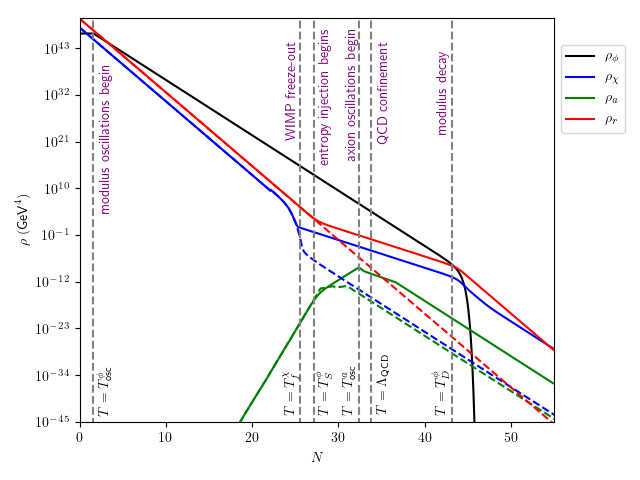}
    \caption{
        Evolution of energy densities $\rho$ versus number of $e$-foldings $N$ in a scenario resembling the annihilation scenario for the WIMP.  Here, we take $m_\phi = 100$ TeV, $m_{\chi}=200$ GeV, $c=27.5$, $B_{\phi \rightarrow \chi} = 0.01$, and $\langle \sigma v \rangle_\chi = 1.05\times 10^{-25} \text{ cm}^3\cdot\text{s}^{-1}$.
        We also take $f_a = 10^{11}$ GeV, $\theta_i = 3.113$, and $T_R = 10^{12}$ GeV.
    }
    \label{fig:energyDensityEvolution_annihilation}
\end{figure}

The remaining parameters in Eqs.~\eqref{rhm}-\eqref{rhr} are then taken as free parameters, yielding the input set 
\begin{equation}
    \{ 
    c, \, 
    m_\phi, \,
    B_{\phi \rightarrow \chi}, \, 
    \langle \sigma v \rangle_\chi, \,
    m_\chi, \, 
    f_a, \, 
    \theta_i, \, 
    T_R
    \}
\end{equation}
where we define $c$ through the expected decay rate formula \cite{Bae:2022okh,Cicoli:2022uqa,Baer:2023bbn}
\begin{equation}
    \Gamma_\phi 
    \simeq
    \frac{
        c
    }{
        48 \pi
    }
    \frac{
        m_\phi^3
    }{
        m_P^2
    }
    .
\end{equation}
The number of relativistic degrees of freedom, $g_*(T)$, is fitted from data used in \cite{Baer:2011uz,Baer:2023bbn}.

\begin{figure}[htb!]
    \centering
    \includegraphics[scale=0.8]{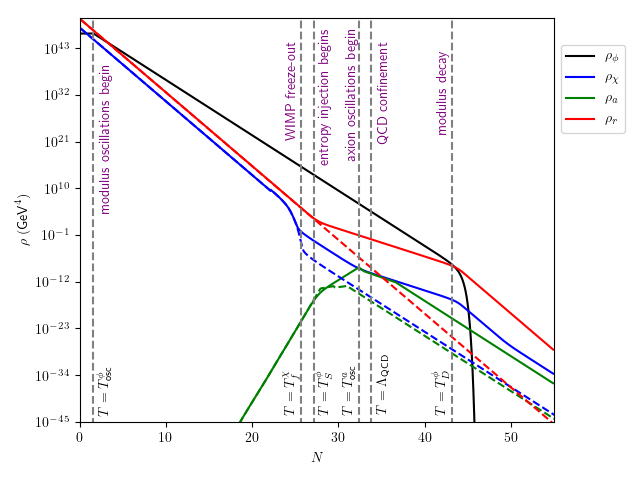}
    \caption{
        Evolution of energy densities $\rho$ versus number of $e$-foldings $N$ in a scenario resembling the branching scenario for the WIMP.  Here, we take $m_\phi = 100$ TeV, $m_{\chi}=200$ GeV, $c=27.5$, $B_{\phi \rightarrow \chi} = 10^{-8}$, and $\langle \sigma v \rangle_\chi = 1.05\times 10^{-25} \text{ cm}^3\cdot\text{s}^{-1}$.
        We also take $f_a = 10^{11}$ GeV, $\theta_i = 3.113$, and $T_R = 10^{12}$ GeV.
    }
    \label{fig:energyDensityEvolution_branching}
\end{figure}

We display the evolution of energy densities using the numeric Boltzmann solutions for parameters that predominantly give the annihilation scenario in Fig.~\ref{fig:energyDensityEvolution_annihilation} and the branching scenario in Fig.~\ref{fig:energyDensityEvolution_branching}.
For both figures, we adopt parameter values motivated by our demands that 1.) the WIMP annihilation cross section is consistent with experimental data, 2.) an EMD phase is prolonged sufficiently to allow axion miniclusters to form based on arguments given by Nelson \textit{et al.} \cite{Nelson:2018via} and Blinov \textit{et al.} \cite{Blinov:2019jqc}, 3.) the modulus decays before BBN, and 4.) the modulus has an effective coupling parameter $c$ that is compatible with explicit models.\footnote{We take $c=27.5$ throughout this work based on results presented in \cite{Cicoli:2022uqa}.  Although it is likely that $c$ could be much larger by the inclusion of additional sectors to which the modulus can decay, our chosen value is near what one might expect to be a lower bound for the effective coupling as it includes only Standard Model degrees of freedom.  As $\Gamma_\phi$ is proportional to $c$, we will see that raising $c$ significantly leads to a larger $T_D^\phi$ and thus makes our results even more constrained.} 
We also display a comparison to the standard thermal evolution (where the modulus is absent), which are shown by the dashed lines of the respective constituents' colors.
The vertical dashed lines in both figures identify qualitative shifts in the cosmology.

\section{Substructure formation} \label{perturb}
To understand the formation of substructure, we need to consider perturbations of the background equations Eqs.~\eqref{back} -- \eqref{rhr}.
%%%%%%%%%%%%%%%%%%%%%%%%%%%%%%%%%%%%%%%%%%%%%%%%%%%
The metric in longitudinal gauge takes the form
\be
ds^2=-\left(1+2 \Phi \right) dt^2 + a(t)^2\left(1-2 \Psi \right) \delta_{ij} dx^i dx^j.
\ee
In the absence of anisotropic stress for the fluid sources, we have $\Phi=\Psi$.
Working in momentum space and suppressing the wave number, the time-time component of the perturbed Einstein equation is
\be 
\label{EE1}
\left( \frac{k^2}{3 a^2 H^2} + 1 \right) \Phi + \Phi' =-\fr{1}{6H^2m_p^{2}}\sum_{\alpha} \delta \rho_{(\alpha)}, 
\ee
where the prime denotes derivatives with respect to number of e-folds $N$, and $v_{(\alpha)}, \delta \rho_{(\alpha)}$, $ \delta p_{(\alpha)}$ are scalar velocity, density and pressure perturbations for each fluid, respectively. Introducing fractional density perturbations $\delta_{(\alpha)} \equiv \delta \rho_{(\alpha)} / \rho_{(\alpha)}$ and defining the velocity perturbation for each fluid as $\theta_{(\alpha)}=a^{-1}\nabla^{2}v_{(\alpha)}$,  the  continuity equations in momentum space are given by
 \bea
\label{ds} \delta'_\phi+ (1+w_\phi) \left(\fr{\ts}{aH}-3\Phi' \right)&=&-\fr{\gam}{H}\Phi \\
\delta'_\chi+ (1+w_\chi)\left(\fr{\tx}{aH}-3\Phi' \right) 
&=& 
B_{\phi \rightarrow \chi} \left(\frac{\Gamma_\phi}{H}\right)\left( \frac{\rhos}{\rhom} \right)\left[ \ds -\dx+ \Phi \right] \nonumber\\ 
&&-\fr{\langle \sigma v \rangle_\chi}{m_\chi H}\left(\rhom  \left(\dx + \Phi\right) -\frac{\overline{\rho}^2_\chi}{\rho_\chi} \left( \Phi + 2 \overline{\delta}_\chi -\delta_\chi \right)\right) \\
\label{dx}
\label{da}
\delta'_a+ (1+w_a)\left(\fr{\tx}{aH}-3\Phi' \right) &=&\frac{dm_a}{dN} \frac{1}{m_a}\Phi - 3c^2_{s,a} \delta_a \\
\delta'_r+\frac{4}{3}\left(\fr{\tr}{aH}-3\Phi' \right)
&=&
(1-B_{\phi \rightarrow \chi}) \left(\fr{\gam}{H}\right) \left( \fr{\rhos}{\rhor} \right)\left[\ds -\dr + \Phi \right] \nonumber \\ 
&+&\fr{\langle \sigma v \rangle_\chi}{m_\chi H}\left(\fr{\rhom^2}{\rhor}\right)\left(\left(2\dx -\dr+ \Phi\right) - \frac{\overline{\rho}_\chi^2}{\rho^2_\chi} \left(2\dx -\dr+ \Phi\right)\right). \;\;\;\;\;
\label{dr} 
\eea
Similarly, the equations for velocity perturbations are
 \bea
\label{ts} \left(1+w_\phi \right)  \left( \ts' + 4\ts-\fr{k^2}{aH}\Phi  \right) -3 \ts  &=& 0 \\
\label{tx}  \left(1+w_\chi \right) \left( \tx' + 4\tx -\fr{k^2}{aH}\Phi \right) - 3 \theta_\chi &=& B_{\phi \rightarrow \chi} \left(\fr{\gam}{H}\right) \left( \fr{\rhos}{\rhom}\right)\left[ \ts -\tx\right] \\
\label{ta}  \left(1+w_a \right) \left( \theta_a^\prime + 4\theta_a -\fr{k^2}{aH}\Phi \right) -3 \theta_a &=& \left( 3 - \frac{dm_a}{dN} \frac{1}{m_a} \right) c^2_{s,a} \theta_a +\frac{k^2}{aH}c_{s,a}^2 \delta_a \;\;\;\;\;\; \\
\tr' -\fr{k^2}{aH}\left(\fr{\dr}{4} +\Phi\right) &=&\left(1- B_{\phi \rightarrow \chi} \right) \left(\fr{\gam}{H}\right) \left( \fr{\rhos}{\rhor}\right) \left[ \frac{3}{4} \ts -\tr\right] \nonumber \;\;\;\;\;\; \\ &&+ \fr{\langle \sigma v \rangle_\chi}{m_\chi H} 
\left( \frac{\rho_\chi^2 - \overline{\rho}_\chi^2 }{\rho_r} \right) \left(\fr{3}{4}\tx-\tr\right).\label{tr}
 \eea
This set of differential equations can be closed by the perturbed Einstein equation Eq.~\eqref{EE1}. 

To calculate the evolution of perturbations, we need to specify initial conditions for the modes.  Following \cite{Fan:2014zua}, we set the modes after modulus domination when all modes of interest are super-Hubble, $k/aH\to 0$. As discussed in \cite{Fan:2014zua}, isocurvature constraints do not apply to the model we are considering here and so we are interested in strictly adiabatic initial conditions for the multi-fluid perturbations
\be\label{aini}
\fr{\delta\rho^{(0)}_{\alpha}}{\rho'_{\alpha}}=\fr{\delta\rho^{(0)}_{\beta}}{\rho'_{\beta}}.
\ee 

We will initially neglect the axion sector as its perturbations decouple from the system above -- we will address the axion perturbations in the next section when discussing miniclusters. 
To establish the initial conditions for the modulus and WIMP sectors, it is useful to find approximate analytic solutions to background well within the modulus dominated phase with $t  \sim H^{-1} \gg  m_\phi^{-1}$. 
An approximate solution is then

\bea 
\label{rhss}\rhos(N) &\simeq & \rho_{\phi}^{(0)}~e^{-3N}\\
\label{rhms}\rhom(N) &=& \rho_{\chi}^{(0)}~e^{-3N/2}\\
\label{rhrs}\rhor(N) &=& \rho_{r}^{(0)}~e^{-3N/2}
\eea  
where we choose initial values so that $ \rho_{\chi}^{(0)},\rho_{r}^{(0)}<<\rho_{\phi}^{(0)}$ and WIMP DM will be primarily of non-thermal origin (i.e., we neglect equilibrium terms). 
The scaling behavior in Eq.~\eqref{rhms} and Eq.~\eqref{rhrs} results from the near cancellation between decays and WIMP annihilation whereas the axion sector decouples and can evolve as matter or radiation. 

Using the background fluid equations with our approximation, and remembering that during modulus domination $\gam /H \ll 1$, it follows that

\be\label{aini2}
\ds^{(0)}=2\dx^{(0)}=2\dr^{(0)}.
\ee
This relation differs from the standard relations due to the presence of decays and entropy production. Taking the super-horizon limit $k/aH\to 0$ of Eq.~\eqref{EE1} in a modulus dominated universe,
we have 
\be\label{EE1SH}
 \Phi\simeq-\fr{1}{6H^2m_p^{2}}\rhos\ds
\ee
where we used conservation of $\Phi$ on super-Hubble scales. Since $\rhos\simeq 3H^2m_p^2$ during modulus domination, Eq.~\eqref{EE1SH} implies the following initial condition for long wavelength gravitational perturbations $\ds^{(0)}=-2\Phi_0$ and it follows from Eq.~\eqref{aini} that 
$\dx^{(0)}=\dr^{(0)}=-\Phi_0$.  Finally, because scalar velocity perturbations quickly decay outside of the Hubble radius their initial values can be ignored. 

\subsection{Perturbation evolution during modulus domination} 
In this section, we examine the evolution of the perturbations for modes that enter the Hubble radius during modulus domination. We note that these modes will be small compared to the size of the horizon at reheating, $k^{-1}< k_{rh}^{-1}$, and thus it will be important for determining the growth of structure at that time.
\\

\noindent{\bf Modulus perturbations}

During modulus domination, we have an effectively matter dominated universe and can therefore set $\Phi=\Phi_0$ for both super and sub-Hubble scales.  Using this, we can rewrite Eq.~\eqref{ts} as 
\be\label{ts*}
\ts' + \ts = \fr{k^2}{H_0}\Phi_0 e^{N/2},
\ee
where we used $H=H_0~e^{-3N/2}$ in a matter dominated Universe.   
This equation can be solved to give the behavior for all wavelengths.
Concentrating on the growing mode, we have    
\be\label{tss}
\ts(k,N)=\fr{2}{3}\fr{k^2}{H_0} \Phi_0 e^{N/2},
\ee
which confirms that long wavelength vector modes are unimportant.
From Eq.~\eqref{tss}, we can derive the evolution of the modulus perturbation $\ds$ during the modulus dominated era subject to the initial condition $\ds^{(0)}=-2\Phi_0$. Noting that until the time of reheating we have $\gam/H \ll 1$ and $\Phi$ is constant, we can rewrite Eq.~\eqref{ds} as
\be\label{dsrw}
\delta'_\phi(k,N)\simeq - \fr{\ts}{H_0}e^{N/2}
\ee
and using the result in Eq.~\eqref{tss}, we integrate to find
\be\label{dss}
%\ds(k,N) =-2\Phi_0  - \fr{2}{3}\fr{k^2}{H_0^2} \Phi_0 \left(  e^{N} \right) - \fr{2}{3}\fr{\Gamma_\sigma}{H_0^2} \Phi_0 \left(  e^{3N/2}  \right) , 
\ds(k,N)\simeq -2\Phi_0 - \fr{2}{3}\fr{k^2}{H_0^2} \Phi_0 e^{N}
\ee
which is again valid on all scales.
\\

\noindent{\bf WIMP and radiation perturbations}

In the absence of source terms for the perturbation equations, the perturbations evolve as expected in a matter dominated universe. However, the additional terms will be important during the period of modulus domination.
Solutions for the complete system can be found as in \cite{Fan:2014zua} where 
by using the background approximation Eqs.~\eqref{rhss} - \eqref{rhrs}, we have time-dependent coefficients scaling as
\bea \label{pref}
 \label{a1} B_\chi\fr{\gam}{H}\left(\frac{\rhos}{\rhom} \right) &\longrightarrow& B_\chi\fr{\gam}{H_0}\left(\frac{\rhos^{(0)}}{\rhom^{(0)}} \right) \equiv A_1 \\
 \label{a2} \fr{\langle \sigma v \rangle}{m_\chi H}\rho_{\chi} &\longrightarrow&  \fr{\langle \sigma v \rangle}{m_\chi H_0}\rho_{\chi}^{(0)} \equiv A_2.
\eea
for density perturbations of the WIMP.
For density perturbations of radiation, from Eq.~\eqref{dr} and Eq.~\eqref{tr} we again find that the scaling cancels and the coefficients are determined by their initial values,
\bea \label{pref2}
 \label{a3} (1-B_\chi)\fr{\gam}{H}\left(\frac{\rhos}{\rhor} \right) &\longrightarrow& (1-B_\chi)\fr{\gam}{H_0}\left(\frac{\rhos^{(0)}}{\rhor^{(0)}} \right) \equiv A_3 \\
 \label{a4} \fr{\langle \sigma v \rangle}{m_\chi H} \left( \frac{\rho_{\chi}}{\rhor} \right){\rho_{\chi}} &\longrightarrow&  \fr{\langle \sigma v \rangle}{m_\chi H_0} \left( \frac{\rho^{(0)}_{\chi}}{\rhor^{(0)}} \right){\rho^{(0)}_{\chi}} \equiv A_4.
\eea
It was shown in \cite{Fan:2014zua} that when selecting a range of initial values motivated from SUSY model building, the annihilation and decay terms are of comparable importance.

The velocity perturbations of the WIMP fluid during modulus domination can be found by using Eq.~\eqref{tss} in Eq.~\eqref{tx}, producing the result \cite{Fan:2014zua}  
\be\label{txs}
\tx(k,N)=\fr{2}{3}\fr{k^2}{H_0} \Phi_0 e^{N/2}.
\ee
Similarly, using Eq.~\eqref{dss} and Eq.~\eqref{txs} in Eq.~\eqref{dx} and remembering that the background coefficients are constants, one finds
\be\label{dxs}
\dx(k,N)=-\Phi_0 - \fr{2}{3} \left( \frac{1+A_1}{1+A_1+A_2} \right) \fr{k^2}{H_0^2} \Phi_0 e^{N}
%\dx(k,N)\simeq-(\Phi_0 + \fr{1}{3}\fr{k^2}{H_0^2} \Phi_0 e^{N}),
\ee  
which again is valid on both super-Hubble and sub-Hubble scales, and we have used the initial conditions $\delta_\phi^{(0)}=2 \delta_\chi^{(0)}=-2 \Phi_0$.

Given the solutions for the modulus and WIMP perturbations, we can solve for the radiation fluid perturbations.
The equation for radiation perturbations is \cite{Fan:2014zua}
\be \label{ode}
\delta_r^{\prime \prime} + \left(2A - \frac{1}{2} \right)\delta_r^\prime+ \left( A^2 -\frac{A}{2} + \frac{k^2}{3 H_0^2} e^N \right)\delta_r= S(N),
\ee
where $A \equiv A_3 + A_4$ and
\be
\alpha = \frac{2}{3} \left( \frac{A_1 A_3 + 2 A_1 A_4 + A_2 A_3 +2 A_4 + A_3}{1+A_1+A_2}\right),
\ee
and the source term is given by
\be \label{sources}
S(N) \equiv -\left(  A^2 -\frac{A}{2} \right) \Phi_0 - \left(  \frac{\alpha}{2}  \left(2A +1 \right) + \frac{2}{3}\left( A+2\right) \right) \frac{k^2}{H_0^2}\Phi_0 e^N .
\ee
A solution to Eq.~\eqref{ode} is
\begin{multline} \label{ode_solution}
\frac{\delta_r}{\Phi_0} = -e^{-AN}\bigg[\cos\bigg\{\fr{2}{\sqrt{3}}\fr{k}{H_0}(e^{N/2} - 1)\bigg\} + \sqrt{3}\frac{AH_0}{k}\sin\bigg\{\fr{2}{\sqrt{3}}\fr{k}{H_0}(e^{N/2} - 1)\bigg\}\bigg] \\ + \sqrt{3}\fr{H_0}{k}e^{-AN}(2A^2 - A)\int_1^{e^{N/2}}x^{2A - 2}\sin\bigg\{\fr{2}{\sqrt{3}}\fr{k}{H_0}(x - e^{N/2})\bigg\}{\rm d}x \\ + \fr{1}{\sqrt{3}}\fr{k}{H_0}e^{-AN}(6A\alpha + 4A + 3\alpha + 8)\int_1^{e^{N/2}}x^{2A}\sin\bigg\{\fr{2}{\sqrt{3}}\fr{k}{H_0}(x - e^{N/2})\bigg\}{\rm d}x, 
\end{multline}
which can be further simplified
by integration by parts.
We thus obtain
\be \label{ode_solution_simplified}
\fr{\delta_r}{\Phi_0} = -1 + \fr{1}{\sqrt{3}}\fr{k}{H_0}[(3\alpha + 2)(2A + 1) + 4]\int_1^{e^{N/2}}x^{2A}\sin\bigg\{\fr{2}{\sqrt{3}}\fr{k}{H_0}(x - e^{N/2})\bigg\}{\rm d}x.
\ee

In the absence of decays and annihilation (corresponding to $A=\alpha=0$ above), the exact solution to Eq.~\eqref{ode} simplifies significantly for all scales:
\be
\delta_r=-4 \Phi_0 + 3 \Phi_0 \cos\left( \frac{2k}{\sqrt{3}H_0} (e^{N/2}-1)\right).
\ee

After passing through the Hubble radius, the radiation modes begin to oscillate with a maximum amplitude reached when the source term in Eq.~\eqref{sources} is in resonance with the oscillations resulting from the homogeneous solution.

There are two important differences when the modulus decay and WIMP annihilation are included in the evolution.
For typical initial values of the radiation and WIMP, as well as decay rates and branching ratios, the oscillations in the modulus dominated phase will be over-damped. A second important effect resulting from decays and annihilations is that these provide an additional source term which acts to boost the amplitude of the density perturbations.  
The enhancement of the amplitude is controlled by the decay and annihilation rates. 
This enhancement results in the saturation of the radiation density perturbation at late times because this growth yields an additional source for the velocity perturbation resulting in a spatial dispersion of the radiation fluid. As the radiation velocity perturbation grows, this slows down the growth of radiation density perturbations. We refer the reader to \cite{Fan:2014zua} for a more detailed analysis. 

\subsection{Evolution through reheating and resulting structure}   

Once modulus decays significantly affect the energy density (and thus the expansion rate), this will lead to `reheating' and alter the evolution of the perturbations. Such effects will become important around an average decay time $t_d \sim H^{-1}_d \sim \Gamma_\phi^{-1}$. 
As the modulus decays become important, the constant scaling from the modulus dominated phase is no longer valid and we will perform the analysis numerically following the approach in \cite{Fan:2014zua}. 

The behavior of the radiation perturbations will change near the time of reheating. Namely, once modulus decay becomes significant it will change the behavior in which they are enhanced as entering the Hubble radius and then saturate due to the balance in the source terms.

The modulus decay will typically happen in less than a Hubble time\footnote{This does not have to be the case, but typically is in models resulting from predictions within particle theory.}. This rapid decay results in a termination of the source terms and the relativistic conversion of modulus particles to radiation acts to wipe out the growth prior to decay.  Thus, the only remnant of the modulus epoch is an extra suppression in the amplitude of radiation perturbations (as a consequence of the decay) \cite{Fan:2014zua,Erickcek:2011us}, which can lead to an enhanced growth of WIMP DM structures on small scales.
The addition of WIMP annihilation as a resulting source for the radiation density does not change this conclusion.  The key effect of adding WIMP annihilation to the system is to effectively introduce an additional source of radiation complementary to that provided by modulus decay,\footnote{This is negligible in the branching scenario.} which increases the rate of initial growth and the importance of the saturation by the source terms. This change also leads to a larger suppression at the time of modulus decay.

For the WIMP perturbations, the density contrast begins to grow linearly with the scale factor upon horizon entry until reheating. The solution Eq.~\eqref{dxs} is an excellent fit to the numerical solution for typical values of $A_1$ and $A_2$ motivated by particle theory \cite{Fan:2014zua}.  After reheating, the WIMP annihilation terms become the dominant sources in Eq.~\eqref{dx} and, along with the radiation production, leads to a power loss in the WIMP density contrast. The annihilation happens in much less than a Hubble time and the resulting density contrast begins to grow logarithmically, as expected in a radiation dominated universe \cite{Brandenberger:1992dw}. In the absence of WIMP annihilation (i.e. in the branching scenario), this result provides the seeds for further growth of structure \cite{Erickcek:2011us}, whereas when annihilation is important they act to further reduce the strength of the perturbations following reheating \cite{Fan:2014zua}.

The WIMP perturbations on scales that enter the Hubble-radius during the EMDE can experience a significant growth. This growth can lead to formation of substructure in the form of compact mini-halos or other objects that provide observations of the non-thermal history \cite{Erickcek:2011us,Fan:2014zua}.
To investigate this possibility one needs to take into account cut-off scales that arise due to kinetic decoupling and free-streaming of DM candidates. We address these questions next.
\\

\noindent{\bf Kinetic decoupling and free-streaming of WIMPs}

Kinetic equilibrium is maintained between the thermal bath and WIMPs through elastic and inelastic scattering processes. Once the rate of these scattering events decreases below the Hubble expansion rate at a temperature $T_{kd}$, this determines the scale $k_{d}$ of kinetic decoupling. 

There are two effects that are important for determining whether the enhanced growth of perturbations from the matter phase will lead to enhanced growth of structure on small scales. The kinetic decoupling scale corresponds to the Hubble scale at the time of decoupling $k^{-1}_{kd} \sim H^{-1}_{kd}$ and at a temperature $T_{kd}$. However, the process is not instantaneous and for temperatures below $T_{kd}$, particles can `free-stream' in and out of over-dense regions in the WIMP distribution. 
This free-streaming scale is determined by an integral involving the average velocity following decoupling until today $t_0$ as
\be
k_{fs}^{-1} = \int_{t_{kd}}^{t_0} \frac{\langle v \rangle}{a} dt.
\ee
Moreover, for scales $k^{-1} < k^{-1}_{kd} = 1/(a_{kd} H_{kd})$, WIMP particles can couple to acoustic oscillations of the thermal bath resulting in further damping on these scales.

To determine which scales maintain the enhancement of the perturbations, it is useful to introduce a Gaussian cutoff into the perturbations
\be
\delta_\chi = \exp \left[ -\frac{k^2}{2k_R^2} \left( \frac{k_R}{k_{cut}}\right)^2 \right] \delta_\chi \left(t_{R}\right)
\ee
where
\be
k_{cut}^{-1} =\mbox{Max}\left( k_{fs}^{-1},k_{kd}^{-1}\right).
\ee
When the modulus decays to WIMPs, the WIMPs will thermalize if their scattering rate exceeds the Hubble rate. In this case, $T_{kd} < T_{D}^\phi$ and $k_{cut}^{-1}=k_{kd}^{-1}$ since $k_{kd}^{-1} > k_{fs}^{-1}$. This leads to the wash-out of the enhanced growth and is typically the case when non-thermal DM thermalizes following reheating. 

Additionally, WIMPs produced from the modulus decay can have significant momentum. This can lead to free-streaming of the dark matter for wavelengths smaller than $k_{fs}^{-1}$. This can also erase the seeded structure resulting from the enhancement during matter domination and suppress structure formation. If WIMPs do not thermalize during reheating, then $T_{kd}>T_D^\phi$ and the WIMPs will decouple prior to reheating. 

There will typically be a continuous momentum spectrum and the average momentum at reheating $\langle p_{R} \rangle$ is model dependent. 
However, we can gain insight by considering when the DM resulting from decay is primarily non-relativistic $\langle p_{R} \rangle \ll m_\chi$. In this case one finds \cite{Fan:2014zua,Erickcek:2011us}
\beq
\frac{k_{R}}{k_{fs}} \approx 
        2\langle v_{R} \rangle  \left(\sinh^{-1}\sqrt{\frac{\sqrt{2}k_{R}}{k_{eq}}}-\sinh^{-1}\sqrt{a_{eq}}\right), 
\eeq
whereas for the relativistic case $ \langle p_{R} \rangle \gg m_\chi$ the result is
\beq
\frac{k_{R}}{k_{fs}} \approx 
        \frac{a_{NR}}{a_{R}}-1 \approx \frac{\langle p_{R} \rangle}{m_\chi}, 
\eeq
where the subscript `$eq$' denotes the time of radiation and matter equality, and 
$a_{NR} \equiv \langle p_{R} \rangle a_{R}/m_\chi$. From the relativistic result, one can see that when the decaying modulus is significantly more massive than the WIMP, free-streaming effects will eliminate the enhanced growth from the modulus epoch.  In the non-relativistic case, however, the reheat scale may be smaller than the free-streaming scale since 
\beq
\frac{k_{rh}}{k_{eq}} = 1.2 \times 10^6 \left(\frac{T_{D}^\phi}{1\,{\rm MeV}}\right) \left(\frac{10.75}{g_{*,s}}\right)^{1/3} \left(\frac{g_*}{10.75}\right)^{1/2}
\eeq
and $\sinh^{-1}x$ behaves as $\log x$ for $x \gg 1$.

%%%%%%%%%%%%%%%%%%%%%%%%%%%%%%%%%%%%%%%%%%%%%%%%%%%

\section{Axion miniclusters and relic density} 
\label{sec:minicluster}

%Intro about Axion, its mass and miniclusters in brief, the background evolution, first order equations governing axion, its analytical solution (from Nelson), conditions of minicluster formation.

In this Section, we discuss the formation of axion miniclusters and expectations for the abundance of axion dark matter in this framework.
As we have previously mentioned, we concentrate on the scenario where PQ breaking occurs before inflation.

The first-order Boltzmann equations for the modulus and axion, including the energy exchange terms, are given by Eqs.~\eqref{ds}, \eqref{da}, \eqref{ts}, and \eqref{ta}. The axion perturbation growth is analogous to the modulus case
Eq.~\eqref{dss}:
\begin{equation}
    \label{eq:perturbationAxionSoln}
    \delta_a(a,k)
    =
    -
    2
    \Phi_0
    -
    \frac{2}{3}
    \left(
        \frac{
            k
        }{
            H(T_{\text{osc}}^a)
        }
    \right)^2
    a
    \Phi_0
\end{equation}
where $H(T_{\text{osc}}^a)$ is the Hubble scale when axion oscillations begin. 
We have also defined our scale factor normalization to be $a(T_{\text{osc}}^a) \equiv 1$. 

In \cite{Nelson:2018via}, a general set of criteria was defined for the formation of axion miniclusters which we follow loosely here.  
Namely, one would expect miniclusters to form efficiently if a perturbation mode grows to $\delta_a \sim 1$ by the end of the EMD phase.  
In this work, we also look at cases where $\delta_a = 10^{-2}$ and $\delta_a = 10^{-3}$ which may also be sufficient for minicluster formation. 
One might expect the minicluster mass $M$ at the time of formation can be estimated by \cite{Nelson:2018via}
\begin{equation}
    \label{eq:miniclusterMass}
    M
    \sim 
    \frac{4}{3}
    \pi 
    k^{-3}
    \, 
    n_a(T)
    \, 
    m_a(T_D^\phi)
\end{equation}
for a mode with wavelength $k^{-1}$ entering the horizon at some temperature $T_{\text{osc}}^a \gtrsim T \gtrsim T_D^\phi $.
For modes which are subhorizon when axion oscillations begin, i.e. for $k > H(T_{\text{osc}}^a)$, only miniclusters with very small masses can form despite the fact that perturbation growth may become large for these modes.
Conversely, for superhorizon modes at the onset of axion oscillations, $k < H(T_{\text{osc}}^a)$, sufficient growth of the perturbations may be difficult.  We therefore focus on the modes which enter the horizon close to $T_{\text{osc}}^a$ so that $k \sim H(T_{\text{osc}}^a)$, which are the modes mostly likely to grow sufficiently to meet the perturbation growth criteria while also producing miniclusters with large masses.  Taking this into consideration, Eq.~\eqref{eq:perturbationAxionSoln} can be rewritten as:
\begin{equation} \label{pert2}
    \delta_a
    \simeq
    -
    2
    \Phi_0
    -
    \frac{2}{3}
    \left( 
        \frac{
            H(T_{\text{osc}}^a)
        }{
            H(T_D^\phi)
        }
    \right)^{2/3}
    \Phi_0
\end{equation}
where $-\Phi_0 \sim 10^{-4}$ and we have the conditions $H(T_D^\phi) \sim \Gamma_\phi \propto (T_D^\phi)^2$ and $H(T_{\text{osc}}^a) \sim m_a(T_{\text{osc}}^a) \propto f_a^{-1}$.
Thus, for a given $T_D^\phi$ we can translate this into a maximal value of the PQ scale $f_a$ which can be expected to form miniclusters with potentially observable masses.

\begin{figure}[htb!]
    \centering
    \includegraphics[scale=0.7]{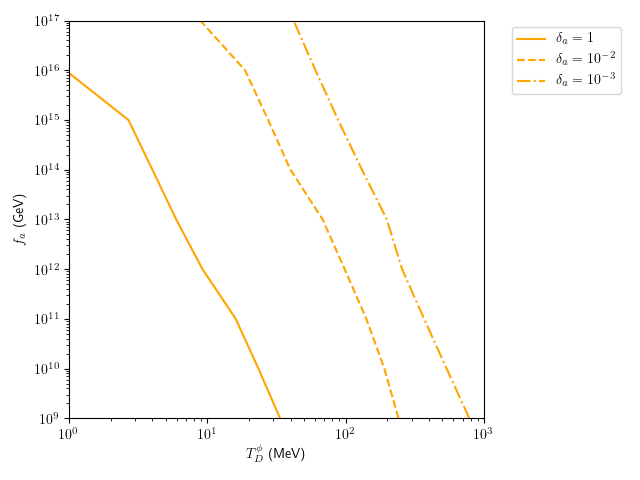}
    \caption{The maximal value of $f_a$ which can be expected to form miniclusters for a given $T_D^\phi$.  The condition for minicluster formation is taken to be  $\delta_a \in \{ 1, 10^{-2}, 10^{-3} \}$ for the solid, dashed, and dot-dashed curves respectively.
    }
    \label{fig:axionMiniclusterBounds}
\end{figure}
We show in Fig.~\ref{fig:axionMiniclusterBounds} this bound on the PQ scale for each $\delta_a \in \{ 1, 10^{-2}, 10^{-3} \}$. 
Here, we note that imposing larger values of  $\delta_a$ for a given $f_a$ also forces $T_D^\phi$ to be closer to the BBN limit.
For example, for $f_a = 10^9$ GeV, $T_D^\phi$ is constrained to be below 33 MeV for $\delta_a = 1$ while it may be as large as 240 MeV for $\delta_a = 10^{-2}$ and 780 MeV for $\delta_a = 10^{-3}$.
These bounds correspond to $m_\phi \lesssim 2.4\times 10^5$ GeV, $m_\phi \lesssim 1.2 \times 10^6$ GeV, and $m_\phi \lesssim 2.8\times 10^6$ GeV, respectively, for our benchmark effective coupling $c=27.5$.
This behavior can be understood from the fact that achieving larger values of $\delta_a$ require longer periods of matter domination, and therefore lower $T_D^\phi$. Conversely, for a given $\delta_a$, increasing $f_a$ (and thus lowering the scale of axion oscillations) requires lowering $T_D^\phi$ for sufficient overlap. The general trends can be understood as follows. From Eq.~\eqref{pert2}, it is clear that  the perturbation growth scales parametrically as  
\be \label{scaling}
\delta_a \, \propto \, \frac{1}{f^{2/3}_a (T_D^{\phi})^{4/3}}.
\ee
Both qualitative behaviors described above can be understood from the parametric scaling in Eq.~\eqref{scaling}.

\begin{figure}[htb!]
    \centering
    \includegraphics[scale=0.7]{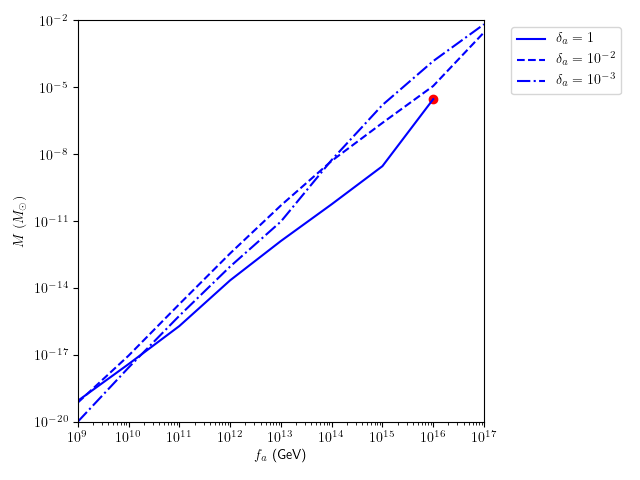}
    \caption{The estimated minicluster mass for modes with $k \sim H(T_{\text{osc}}^a)$ at the time of formation, $T_D^\phi$, for a given $f_a$.
    Superhorizion modes which have $k \gtrsim H(T_{\text{osc}}^a)$ may produce larger miniclusters than those displayed.
    The condition for minicluster formation is taken to be  $\delta_a \in \{ 1, 10^{-2}, 10^{-3} \}$ for the solid, dashed, and dot-dashed curves respectively.
    }
    \label{fig:axionMiniclusterMass}
\end{figure}

We can also estimate the mass of the miniclusters at the time of formation for these modes.  
As we are primarily focused on modes entering the horizon close to $T_{\text{osc}}^a$, Eq.~\eqref{eq:miniclusterMass} can be rewritten as 
\begin{equation}
    \label{eq:miniclusterMass2}
    M
    \sim 
    \frac{2}{3}
    \pi 
    \left( 
        \frac{
            m_a(T_D^\phi)
        }{
            m_a(T_{\text{osc}}^a)
        }
    \right)
    \frac{
        \mathcal{A}_0^2
    }{
        m_a(T_{\text{osc}}^a)
    }
\end{equation}
where $\mathcal{A}_0^2$ is given by Eq.~\eqref{eq:axionAmplitude}.
We show this estimate in Fig.~\ref{fig:axionMiniclusterMass}.
Here, we adopt $T_D^\phi$ and corresponding $T_{\text{osc}}^a$ values from the bounds on $(T_D^\phi, f_a)$ pairs for the various $\delta_a$ which are shown in Fig.~\ref{fig:axionMiniclusterBounds}.
The differences between the three curves shown in Fig.~\ref{fig:axionMiniclusterMass} all stem from variations of $m(T_D^\phi)$ and $m(T_{\text{osc}}^a)$ along these bounds.
For values of the PQ scale $f_a \lesssim \mathcal{O}(10^{14} \text{ GeV})$ and $f_a \lesssim \mathcal{O}(10^{10} \text{ GeV})$ for $\delta_a = 10^{-3}$ and $\delta_a = 10^{-2}$, respectively, we have the maximally allowed $T_D^\phi \gtrsim \Lambda_{\text{QCD}}$ which causes a reduction in $m_a(T_D^\phi)$ compared to values which have $T_D^\phi \lesssim \Lambda_{\text{QCD}}$.
Conversely, the larger values of $T_D^\phi$ allowed by smaller $\delta_a$ require an increase the axion oscillation temperature, as can be seen from Eq.~\eqref{eq:axionOscillationHubble}.
This causes a reduction in $m_a (T_{\text{osc}}^a)$ which, as can be seen in the figure, can compete against the previously mentioned reduction in $m_a(T_D^\phi)$.
We also note that for $\delta_a = 1$, we do not display $f_a \gtrsim 10^{16}$ GeV as the minicluster formation criteria can only be met for $T_D^\phi$ in violation of BBN bounds.

Finally, it is worth noting that the estimated minicluster masses in Fig.~\ref{fig:axionMiniclusterMass} are not necessarily the maximum minicluster masses that one could predict in this framework.
By loosening our demands on the perturbation growth $\delta_a$, it is possible that much larger superhorizon modes $k < H(T_{\text{osc}}^a)$ may meet these demands and thus produce miniclusters with even larger masses.
We leave a more precise study of minicluster masses with weakened perturbation growth constraints and the study of their evolution to the present day for future work.

\begin{figure}[htb!]
    \centering
    \includegraphics[scale=0.7]{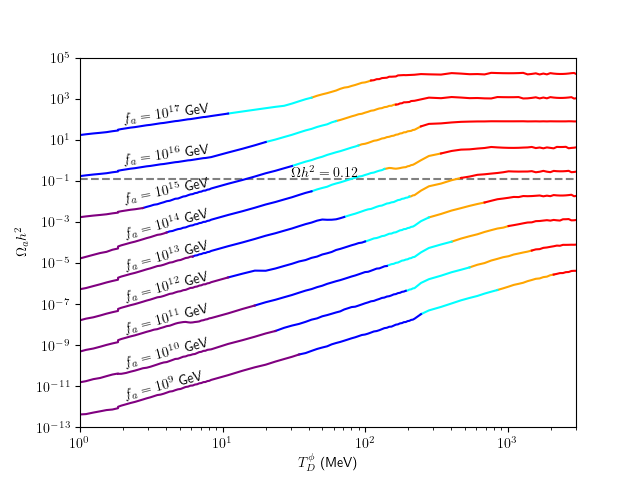}
    \caption{
        Axion relic density versus $T_D^\phi$ where $\theta_i = 2$. 
        For our benchmark, these values of $T_D^\phi$ correspond to $m_\phi \in [ 2.4\times 10^4, \, 5 \times 10^6 ]$ GeV. Purple regions are consistent with $\delta_a = 1$, dark blue with $\delta_a = 10^{-2}$, and light blue with $\delta_a = 10^{-3}$. Orange regions correspond to even lower values of $\delta_a$ but still have some overlap with an EMDE.
        Red regions have no overlap with an EMDE.
    }
    \label{fig:axionDM_untuned}
\end{figure}

\begin{figure}[htb!]
    \centering
    \includegraphics[scale=0.7]{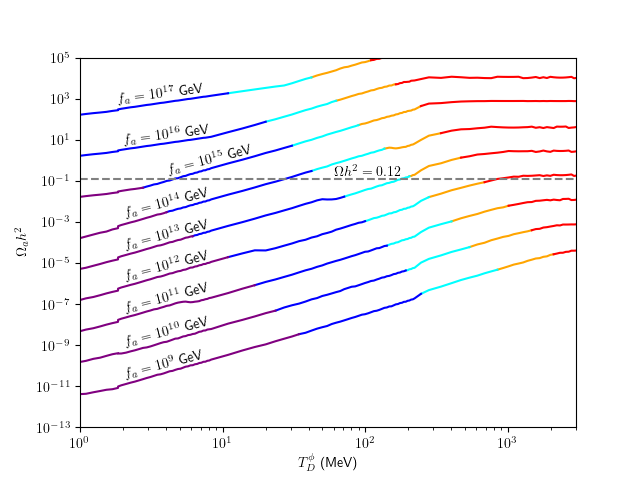}
    \caption{
        Axion relic density versus $T_D^\phi$ where $\theta_i = 3.113$. 
        For our benchmark, these values of $T_D^\phi$ correspond to $m_\phi \in [ 2.4\times 10^4, \, 5 \times 10^6 ]$ GeV. Purple regions are consistent with $\delta_a = 1$, dark blue with $\delta_a = 10^{-2}$, and light blue with $\delta_a = 10^{-3}$. Orange regions correspond to even lower values of $\delta_a$ but still have some overlap with an EMDE.
        Red regions have no overlap with an EMDE.
    }
    \label{fig:axionDM_tuned}
\end{figure}

Now that we have an understanding of the regions of parameter space where axion miniclusters may be expected to form, we can look at the expected relic abundance of axion dark matter within these regions.
In Fig.~\ref{fig:axionDM_untuned}, we display the axion relic density computed from numerical solutions to the Boltzmann equations as a function of $T_D^\phi$ for an untuned $\theta_i = 2$.  
Purple regions are consistent with the minicluster formation criterion $\delta_a = 1$, dark blue with $\delta_a = 10^{-2}$, and light blue with $\delta_a = 10^{-3}$. 
Orange regions correspond to even lower values of $\delta_a$  but still have some overlap with an EMDE.
Red regions, however, have $T_{\text{osc}}^a < T_D^{\phi}$ and thus represent axions which only begin oscillating after the EMDE phase has ended. 
From the figure, we also see the red regions rapidly approach constant values; this is as expected since for $T_D^\phi > T_{\text{osc}}^a$, the axion is still frozen until any entropy release from modulus decay has ceased.
However, any overlap with an EMDE then necessarily implies overlap with the era of entropy release, resulting in a reduction of the axion yield $Y_a$ and thus a reduction in its relic abundance.

It is immediately evident from Fig.~\ref{fig:axionDM_untuned} that, at least for EMDEs consistent with our demand of minicluster formation, saturating the total DM relic density with axions is only viable for $f_a \gtrsim \mathcal{O}(10^{13})$ GeV.  
For lower values of $f_a$, the horizontal dashed line displaying $\Omega h^2 = 0.12$ can only intersect with red regions. 
Additionally, if we impose the demand for both minicluster formation and a DM relic density which is saturated by axions, we are forced to have $T_D^{\phi} \, \lesssim \mathcal{O}(100)$ MeV. 
As expected, the regions compatible with minicluster formation become smaller the larger $\delta_a$ is. For $\delta_a = 1$, for example, we are forced into the purple regions which only intersect the relic density line when $\phi$ decays very close to BBN and $f_a \sim \mathcal{O}(10^{16})$ GeV.  Additionally, if $T_D^\phi$ is further required to be above $\sim 4$ MeV for compatibility with BBN as suggested in \cite{Kawasaki:2000en,Hasegawa:2019jsa}, our results become even more constrained so that only $f_a \lesssim \mathcal{O}(10^{14})$ GeV is allowed and the total relic density cannot be satisfied with the choice $\theta_i=2$ for the misalignment angle.

The general trends can once again be understood from the parametric scaling in Eq.~\eqref{scaling}. For a given value of $f_a$, increasing $T_D^{\phi}$ means shorter periods of matter domination and hence smaller values of $\delta_a$; thus, purple regions transition to dark blue, then light blue, and finally orange. On the other hand, for a given $T_D^{\phi}$ increasing $f_a$ also lowers  $\delta_a$, causing the viable regions for minicluster formation to shrink as $f_a$ increases.

For values of the misalignment angle which approach the limiting value, $\theta_i \rightarrow \pi$, the axion relic abundance can be enhanced by a potentially large amount due to anharmonic effects \cite{Visinelli:2009zm}. 
We display the axion relic density for a tuned $\theta_i = 3.113$ in Fig.~\ref{fig:axionDM_tuned}, again computed from numerical solutions to the Boltzmann equations.
The color coding is identical to Fig.~\ref{fig:axionDM_untuned}. 
Generally, the trends discussed in the case of $\theta_i = 2$ continue to hold, with the contours of constant $f_a$ pushed upwards to higher relic densities. 
It is clear that, even with a mild amount of tuning, demanding both minicluster formation and purely axion DM is only viable for  low $T_D^\phi \lesssim \mathcal{O}(200)$ MeV and $f_a \sim \mathcal{O}(10^{13})$ GeV. 
Meeting both demands for lower values of the PQ scale $f_a$ thus clearly requires a large degree of tuning for the misalignment angle.

\section{Relic density contributions from WIMPs}
\label{sec:WIMPs}

In light of the results in Section \ref{sec:minicluster}, we might expect that if axion miniclusters do form, there should be additional contributions from other species to the total amount of dark matter. 
The parameter space with $\delta_a \sim\mathcal{O}(10^{-3} - 1)$  for values of $f_a \lesssim \mathcal{O}(10^{14})$ GeV typically has axions contributing as a sub-dominant component of dark matter, as seen in both Fig.~\ref{fig:axionDM_untuned} for an untuned case and in Fig.~\ref{fig:axionDM_tuned} for a mildly tuned case. 
Demanding values of $\delta_a$ at the larger end of this range forces one into larger values of $f_a$ if axions are to constitute a significant percentage of the total dark matter. Indeed, if one requires $\delta_a \sim 1$, one ultimately runs out of parameter space even for large $f_a$, as the BBN limit on $T_D^{\phi}$ is breached. 
 
Since it appears that axion-only dark matter is  unlikely to saturate the observed dark matter density while also producing miniclusters, we are led to consider a multi-component dark matter scenario.  
An appreciable amount - if not a majority - of the dark matter might then be composed of some other dark matter particle, which allows for the formation of axion miniclusters while easing the requirement of extreme misalignment angle tuning.  In this work, we study the case where the second component is composed of WIMPs. 

Generally, one would expect that the WIMP annihilation cross section can be well-approximated by $s$-wave and $p$-wave contributions:
\begin{equation}
    \langle 
    \sigma v
    \rangle_{\chi} 
    \sim 
    s_0
    +
    p_0 
    \frac{T}{m_\chi}
    .
\end{equation}
From the WIMP Boltzmann equation given in Eq.~\eqref{rhm}, we can see that while WIMPs are relativistic, the annihilation cross section primarily serves to maintain equilibrium.  However, once $T \lesssim m_\chi$ and the equilibrium density $\overline{\rho}_\chi$ becomes Boltzmann-suppressed, the precise value of $\langle \sigma v \rangle_\chi$ plays a large role in determining the final WIMP relic density.  Thus, the $p$-wave contributions are only very important for our considerations if $p_0 \gtrsim s_0$.
Here we focus on the case of a purely $s$-wave annihilation cross section so that $\langle \sigma v \rangle_\chi = s_0$ and leave the analysis of the $p$-wave contributions for future work.

WIMP annihilations can produce detectable gamma-ray signals.  Such signals have been the focus of many experiments such as HESS \cite{HESS:2016mib}, MAGIC \cite{MAGIC:2016xys}, and Fermi-LAT \cite{Fermi-LAT:2015att,Fermi-LAT:2016uux}, which have constrained the $T \rightarrow 0$ limit of $\langle \sigma v \rangle_\chi$ for several different production channels.  Here we take a model-independent approach and adopt the limits from the $\tau^+ \tau^-$ channel shown in \cite{Fermi-LAT:2015att}.  This particular channel tends to be an upper bound on the WIMP annihilation cross section.  Other channels (if relevant for a given model) will then be even more highly constrained.  As we will see, even under the most optimistic constraints it is \textit{extremely difficult} to produce reasonable amounts of WIMP dark matter in the regime where axion miniclusters may be expected to form.

\begin{figure}[htb!]
    \centering
    \includegraphics[scale=0.7]{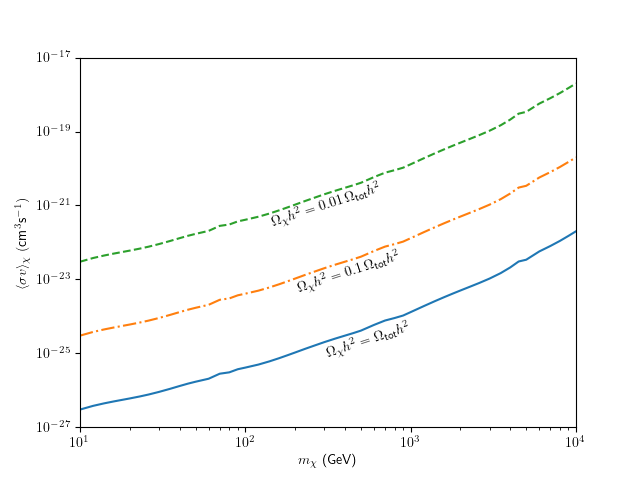}
    \caption{Upper limits on the annihilation of WIMP DM to  $\tau^+ \tau^-$ taken from Fermi-LAT \cite{Fermi-LAT:2015att}. Here, we rescale the constraints assuming that the 
     observed DM density is composed of 100\% WIMPs (blue    curve), 10\% WIMPs (orange curve), and 1\% WIMPs (green curve).
    }
    \label{fig:crossSections}
\end{figure}

Firstly, the Fermi-LAT upper bounds on the annihilation cross section to $\tau^+ \tau^-$ needs to be reinterpreted in the context of a  multi-component dark matter scenario. The constraints become weaker if WIMPs make up a smaller percentage of the total dark matter density; specifically, the WIMP constraint 
$
    \langle 
    \sigma
    v
    \rangle_\chi^{\text{multi}}
$
in a multi-component scenario can be related to the constraint
$
    \langle 
    \sigma
    v
    \rangle_\chi^{\text{single}}
$
in a WIMP-only scenario by \cite{Baer:2016ucr}
\begin{equation}
    \langle 
    \sigma
    v
    \rangle_\chi^{\text{multi}}
    =
    \xi^2
    \langle 
    \sigma
    v
    \rangle_\chi^{\text{single}}    
\end{equation}
where $\xi \leq 1$ is the percentage of total dark matter being composed of WIMPs. In Fig.~\ref{fig:crossSections}, we show the upper limits on the annihilation of WIMPs  from Fermi-LAT, appropriately rescaled assuming that the   observed DM density is composed of 100\% WIMPs (blue    curve), 10\% WIMPs (orange curve), and 1\% WIMPs (green curve).

\begin{figure}[htb!]
    \centering
    \includegraphics[scale=0.7]{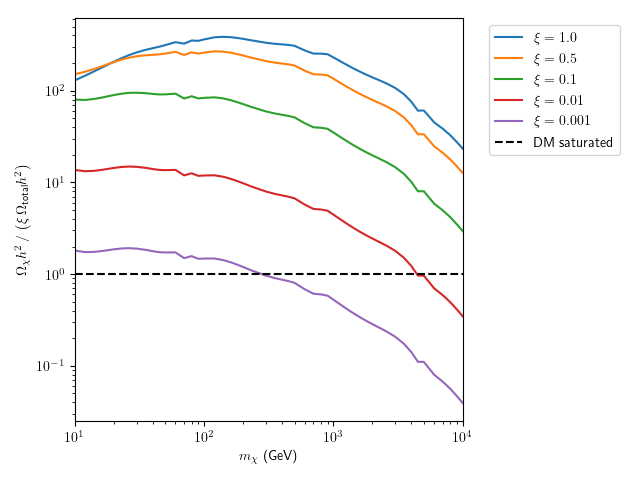}
    \caption{
        Annihilation Scenario: Normalized WIMP relic density $\Omega_\chi h^2$ versus WIMP mass $m_\chi$ for $m_\phi = 4.1 \times 10^{5}$ GeV ($T^\phi_D =  51$ MeV).
        We display curves for values of $\xi \in \{ 0.001, \, 0.01, \, 0.1, \, 0.5, \, 1 \}$ where $\xi$ is the percentage of total dark matter composed of WIMPs.
        We fix the annihilation cross section $\langle \sigma v \rangle_\chi^{\text{multi}}$ to the maximally allowed value by Fermi-LAT constraints, $\xi^2 \langle \sigma v \rangle_\chi^{\text{single}}$, as in Fig.~\ref{fig:crossSections}.
        We have also fixed $B_{\phi \rightarrow \chi} = 5\%$.
    }
    \label{fig:wimpAnnihilation}
\end{figure}

\begin{figure}[htb!]
    \centering
    \includegraphics[scale=0.8]{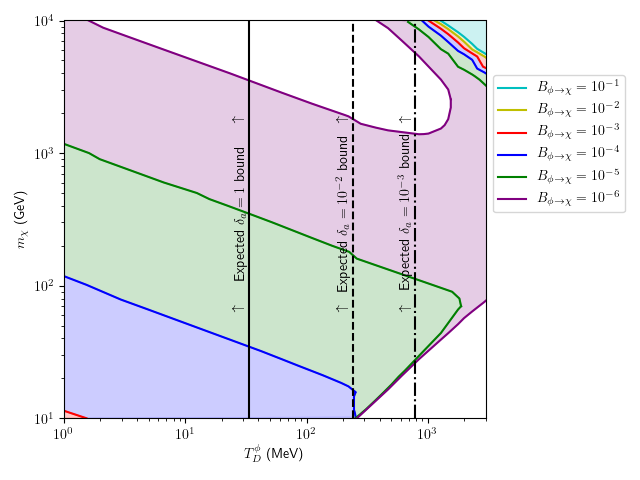}
    \caption{
        Branching Scenario: We display  contours of  the branching of the modulus to WIMPs $B_{\phi\rightarrow\chi}$ for which the observed DM relic density is obtained, assuming Fermi-LAT constraints with $\xi=1$. 
        Here, the solid boundaries of contours show where the overall DM relic density is saturated exclusively by WIMPs, while shaded regions have $\Omega_\chi h^2 < 0.12$.
        For $T_D^\phi \gtrsim 0.2$ GeV, the scenario transitions to the annihilation scenario for the values of $B_{\phi \rightarrow \chi}$ shown.
        The black vertical lines indicate the highest $T^\phi_D$ for which axion minicluster formation is expected to occur for $\delta_a \in \{ 1, \, 10^{-2}, \, 10^{-3} \}$. 
    }
    \label{fig:wimpContours}
\end{figure}

We now discuss the confluence of the axion perturbation condition, relic density, and Fermi-LAT upper limits on the WIMP annihilation cross section. From Sec.~\ref{sec:minicluster} we see that the axion perturbation condition prefers low $T^\phi_D \, \sim \, \mathcal{O}(1-10)$ MeV. We first understand the implications of this for WIMPs in the annihilation scenario, where the relic density is given by Eq.~\eqref{relicannnont}. For low $T^\phi_D$ and $T_f \, \sim \, m_\chi/20 \, \sim \, 5$ GeV (for a 100 GeV WIMP), we have 
\be \label{factoranni}
T_f/T^\phi_D \, \sim \,  \mathcal{O}(100)
\ee
which implies that the annihilation scenario will generally oversaturate the relic density if the condition for axion miniclusters is satisfied. 
To compensate for the large factor in Eq.~\eqref{factoranni} we consider WIMP candidates with highly enhanced annihilation cross sections; however, that option is subject to upper limits from Fermi-LAT. 

This situation is portrayed in Fig.~\ref{fig:wimpAnnihilation}. We take $B_{\phi\rightarrow\chi}=5\%$ so that we are indeed in the annihilation scenario.  
We also set $m_\phi = 4.1 \times 10^5$ GeV so that $T^\phi_D \simeq 51$ MeV, roughly corresponding to the upper limit allowed by $\delta_a = 1$.
The vertical axis is the normalized WIMP contribution to  the DM relic density, which should be - \textit{at most} - unity (shown by the black dashed line) for consistency. 
The blue curve depicts the normalized WIMP contribution to the relic density corresponding to  the Fermi-LAT upper limits on the annihilation cross section when WIMPs comprise 100\% of DM. 
Clearly, WIMPs oversaturate the relic density in this case, unless they are extremely heavy and above the reach of Fermi-LAT. 
Reducing the WIMP contribution (reducing $\xi$) mitigates the tension somewhat, but is generally unable to overcome the effect of the large value of $T_f/T^\phi_D$. We therefore see that if the annihilation cross section must be raised significantly to produce a sensible quantity of WIMPs, we are forced to consider scenarios where WIMPs make up a small percentage of the total dark matter.  
We are therefore again plagued by the tuning problems of our axion-only DM scenario of the previous section unless yet another DM candidate is considered.

We now move on to the branching scenario for WIMPs. In this case, the WIMP yield is given by Eq.~\eqref{branchingyield}. We first describe the parametric scaling of the DM yield and modulus branching, and Fig.~\ref{fig:wimpContours} gives our quantitative results.
Firstly, we note that the modulus yield scales parametrically as
\be
Y_\phi \, \sim \, T^\phi_D/m_\phi \, \propto \, (T^\phi_D)^{1/3}\,\,.
\ee
Comparing to dark matter observations implies a WIMP yield of 
\begin{equation}
    Y_\chi 
    \simeq
    5
    \times 10^{-10}
    \left(
        \frac{1 \text{ GeV}}{m_\chi}
    \right)
\end{equation}
and so the branching ratio is
\be
B_{\phi \rightarrow \chi}
\sim 
\mathcal{O}
\left(
10^{-4}
\right)
\times
\left(
    \frac{10 \text{ GeV}}{m_\chi}
\right)
\left(
    \frac{m_\phi }{10^5 \text{ GeV}}
\right)
\left(
    \frac{0.05 \text{ GeV}}{T_D^\phi}
\right).
\ee
Thus, for $T^\phi_D \sim \mathcal{O}(1-50)$ MeV required for axion minicluster formation the branching of the modulus to WIMPs falls in the range $B_{\phi\rightarrow\chi} \sim 10^{-4} - 10^{-7}$ depending on the precise magnitude of $m_\chi$. 
This is shown in Fig.~\ref{fig:wimpContours}.
The larger branching ratios are concentrated on the top right corner, where the annihilation scenario is operational. 
For the smaller branchings displayed $B_{\phi\rightarrow\chi} \sim 10^{-4} - 10^{-6}$, the annihilation scenario is also operational beginning at $T_D^\phi \gtrsim 0.2$ GeV for small WIMP masses.
This results from $m_\chi$ being sufficiently low so that the late WIMP freeze-out temperature (close to $T_D^\phi$) results in $\overline{\rho}_\chi$ being large enough to produce WIMPs through thermal interactions. 
For larger $m_\chi$, the WIMP freeze-out temperature is then larger than $T_D^\phi$ so these reactions do not occur and the branching scenario takes over.

In Fig.~\ref{fig:wimpContours}, the vertical dashed lines show the maximal value of $T^\phi_D$ that is compatible with axion minicluster formation, as obtained from Fig.~\ref{fig:axionDM_untuned}.
We see that for the $\delta_a = 1$ bound, the total DM relic density can indeed be saturated for the displayed branching ratios $B_{\phi \rightarrow \chi} \sim 10^{-4} - 10^{-6}$ for most of the range of $m_\chi$ probed by Fermi-LAT.
However as $\delta_a$ decreases the parameter space becomes more restricted, requiring heavier WIMPs with $m_\chi \gtrsim 1-2$ TeV to have $B_{\phi \rightarrow \chi} \lesssim 10^{-7}$. 
Many currently existing models involving moduli and WIMPs predict $B_{\phi \rightarrow \chi} \sim \mathcal{O}(0.01-0.1)$, and are thus in tension with these results - although this certainly does not preclude the possibility of a compatible WIMP model with such a low $B_{\phi \rightarrow \chi}$.
We reiterate that here we take an agnostic approach as to the model-building details and simply regard the branching ratio as a free parameter.

Based on these results, it appears that the branching scenario - which is capable of saturating the total DM relic density - may be preferable in regions of parameter space that allow for axion minicluster formation within a multicomponent WIMP and axion dark matter framework.
It is interesting to note that, as discussed in Sec.~\ref{perturb}, the branching scenario may allow for the sustained growth of WIMP DM substructures while the annihilation scenario typically has any growth washed out by rapid thermalization.
Thus, unless these effects are washed out by free-streaming (as may be likely for small WIMP masses), it appears plausible that WIMP substructures are likely form in tandem with axion miniclusters within this framework.
We leave a thorough analysis of these substructures for future work.

\section{Conclusions}

In this paper, we have studied the question of DM substructure formation in the presence of an EMDE, with a focus on axion miniclusters. 
We have especially dealt with the case of PQ breaking before inflation, where EMDEs can play a critical role in ensuring minicluster formation. Our results are as follows.

Firstly, the axion oscillation temperature must be larger than the modulus decay temperature: $T^a_{\rm osc} \, > \, T^\phi_D$. 
Secondly, the axion minicluster formation condition Eq.~\eqref{pert2} requires $H(T_{\text{osc}}^a)  \, \gg \, H(T_D^\phi)$. 
However, since $H(T_D^\phi) \sim \Gamma_\phi \propto (T_D^\phi)^2$ and $H(T_{\text{osc}}^a) \sim m_a(T_{\text{osc}}^a) \propto f_a^{-1}$, these conditions effectively force $T^\phi_D$ to be near BBN, and especially for cases when the axion constitutes a significant part of the relic density.  
The effect of demanding axion minicluster on WIMPs then is that they are forced into the branching scenario, in order to be compatible with upper limits coming from Fermi-LAT.
If WIMPs comprise a significant percentage of the total dark matter so the constraints on the PQ sector are eased, a branching ratio $B_{\phi \rightarrow \chi} \lesssim \mathcal{O}(10^{-4})$ is then required so that the total relic density does not become oversaturated. 
Furthermore, because the branching scenario is required in this regime, our results suggest that models which can meet the demands on the PQ sector and on $B_{\phi \rightarrow \chi}$ for minicluster formation may also predict WIMP DM substructures - especially for heavier WIMPs - unless free-streaming effects become significant.

Despite the effective decoupling of WIMPs and axions, the level of constraint on one sector by imposing demands on the other within our EMDE framework is quite remarkable.  
These conclusions also pave several avenues for future work.
In particular, a detailed study of free-streaming effects for WIMPs produced through the branching scenario would reveal to what degree WIMP substructure formation is indeed correlated with axion substructure formation.
Additionally, the inclusion of sizeable $p$-wave contributions to the WIMP cross section may change these predictions as 1.) the WIMP relic density may be decreased by a large degree in the annihilation scenario while the $T\rightarrow 0 $ limit on $\langle \sigma v \rangle_\chi$ remains compatible with Fermi-LAT bounds, and 2.) this may result in a larger degree of WIMP thermalization in the branching scenario, which likely erases any tentative WIMP substructure growth.
Finally, it is interesting to question whether other dark matter candidates such as Primordial Black Holes could share such a high degree of constraint in the context of a similar EMDE multicomponent DM model.

\section*{Acknowledgements}
We thank Amy Burks, Adrienne Erickeck, and Nemanja Kaloper for useful conversations.  S.W. thanks KITP Santa Barbara and the Simons Center for hospitality. This research was supported in part by DOE grant DE-FG02-85ER40237. K.S. thanks the Simons Center for hospitality and is supported in part by DOE grant DE-SC0009956.
The computing for this project was performed at the OU Supercomputing Center for Education \& Research (OSCER) at the University of Oklahoma (OU).

%\bibstyle{jhep}
\bibliography{ref}

%\printbibliography %Prints bibliography

\end{document}